\documentclass[10pt,superscriptaddress,aps,twocolumn,secnumarabic,nobibnotes]{revtex4-1}

\usepackage{fullpage}
\usepackage{url}
\usepackage{amssymb}
\usepackage{amsmath}
\usepackage{bm}
\usepackage[a]{esvect}
\usepackage{braket}
\usepackage{paralist}
\usepackage[colorlinks=false, pdfborder={0 0 0}]{hyperref}
\usepackage{bm}
\usepackage{listings}
\usepackage{xcolor}
\usepackage{subfigure}
\usepackage{graphicx}

\newcommand{\code}[1]{\texttt{#1}} 
\newcommand{\du}{\ensuremath\mathrm{d}}
\newcommand{\iu}{\ensuremath\mathrm{i}}
\newcommand{\Matlab}{\textsc{Matlab}}
\newcommand{\beq}{\begin{equation}}
\newcommand{\eeq}{\end{equation}}
\newcommand{\e}{\ensuremath\mathrm{e}} 
\newcommand*{\ped}[1]{\ensuremath{_\mathrm{#1}}} 
\renewcommand*{\ap}[1]{\ensuremath{^\mathrm{#1}}}

\newcounter{bla}

\begin{document}

\title{\code{CPMC-Lab}: A \Matlab{} Package for \\ Constrained Path Monte Carlo Calculations}
\date{July 29, 2014}

\author{Huy Nguyen}
\affiliation{Department of Physics, Reed College, Portland, OR 97202, USA}
\affiliation{Department of Physics, College of William and Mary, Williamsburg, VA 23185, USA}
\author{Hao Shi}
\author{Jie Xu}
\author{Shiwei Zhang}
\affiliation{Department of Physics, College of William and Mary, Williamsburg, VA 23185, USA}

\begin{abstract}
We describe \code{CPMC-Lab}, a \Matlab{} program for the constrained-path and phaseless auxiliary-field Monte Carlo methods. These methods have allowed applications ranging from the study of strongly correlated models, such as the Hubbard model, to \emph{ab initio} calculations in molecules and solids. The present package implements the full ground-state constrained-path Monte Carlo (CPMC) method in  \Matlab{} with a graphical interface, using the Hubbard model as an example. The package can perform calculations in finite supercells in any dimensions, under periodic or twist boundary conditions. Importance sampling and all other algorithmic details of a total energy calculation are included and illustrated. This open-source tool allows users to experiment with various model and run parameters and visualize the results. It  provides a direct and interactive environment to learn the method and study the code with minimal overhead for setup. Furthermore, the package can be easily generalized  
for auxiliary-field quantum Monte Carlo (AFQMC) calculations in many other models for correlated electron systems, and can serve as a template for developing a production code for AFQMC total energy calculations in real materials. Several illustrative studies are carried out in one- and two-dimensional lattices on total energy, kinetic energy, potential energy, and charge- and spin-gaps.

\end{abstract}

\maketitle

{\bf PROGRAM SUMMARY}

\begin{small}
\noindent
{\em Manuscript Title:} \code{CPMC-Lab}: A \Matlab{} Package for Constrained Path Monte Carlo Calculations\\
{\em Authors:} Huy Nguyen, Hao Shi, Jie Xu and Shiwei Zhang                                                \\
{\em Program Title:} \texttt{CPMC-Lab}                                          \\
{\em Journal Reference:}                                      \\
{\em Catalogue identifier:}                                   \\
{\em Licensing provisions:} CPC non-profit license \\
{\em Programming language:} \Matlab                              \\
{\em Computer:}  The non-interactive scripts can be executed on any computer capable of running \Matlab{} with all \Matlab{} versions. The GUI requires \Matlab{} R2010b (version 7.11) and above.                                            \\
{\em Operating system:} Windows, Mac OS X, Linux                                     \\
{\em RAM:} variable.                                              \\
{\em Number of processors used:} 1                              
{\em Classification:} 7.3 Electronic Structure                                         \\
{\em External routines/libraries:} \Matlab                            \\
{\em Nature of problem:} Obtaining ground state energy of a repulsive Hubbard model in a supercell in any number of dimensions\\
{\em Solution method:} In the Constrained Path Monte Carlo (CPMC) method, the ground state of a many-fermion system is projected from an initial trial wave function by a branching random walk in an overcomplete basis of Slater determinants. Constraining the determinants according to a trial wave function $\ket{\Psi\ped{T}}$ removes the exponential decay of the signal-to-noise ratio characteristic of the sign problem. The method is exact if $\ket{\Psi\ped{T}}$ is exact. \\
{\em Unusual features:} Direct and interactive environment with a Graphical User Interface for beginners to learn and study the Constrained Path Monte Carlo method with minimal overhead for setup. 

\vskip 2 em

\textbf{References}

\vskip 1 em

[1] S.~Zhang, J.~Carlson and J.~E.~Gubernatis, Phys.\ Rev.\ B.\ \textbf{55}, 7464 (1997).

[2] Shiwei Zhang, in the free online book ``Emergent Phenomena in Correlated Matter Modeling and Simulation, vol 3,'' eds.~E. Pavarini, E. Koch, and U. Schollwock (2013). \url{http://www.cond-mat.de/events/correl13}
\end{small}

\section{Introduction}
\label{sec: intro}

The study of interacting quantum many-body systems remains an outstanding challenge, especially systems with strong particle interactions, where perturbative approaches are often ineffective. Numerical simulations provide a promising approach for studying such systems. One of the most general numerical approaches is quantum Monte Carlo (QMC) methods based on auxiliary fields, which are applied in condensed matter physics, nuclear physics, high-energy physics and quantum chemistry. These methods allow essentially exact calculations of ground-state and finite-temperature equilibrium properties of interacting many fermion systems. 

As is well-known, however, these methods suffer from the sign problem which severely limits their applicability. Considerable progress has been achieved in circumventing this problem by constraining the random walks while sampling the space of auxiliary fields~\cite{Zhang2013-Juelich-LectNotes}. Many applications of this method involve lattices where there is a sign problem, for example, Hubbard-like models where the local interactions lead to auxiliary-fields that are real. In these cases the method is known as Constrained Path Monte Carlo (CPMC)~\cite{cpmc0K-2,PhysRevB.78.165101}. The method can also be generalized to treat realistic electron-electron  interactions to allow for \emph{ab initio} calculations on real materials~\cite{PhysRevLett.90.136401,al-saidi:224101}. For these systems there is a phase problem because the Coulomb interaction leads to complex auxiliary fields. In such systems, the method is referred to as phaseless or phase-free auxiliary-field QMC (AFQMC)\@.
In both cases (lattice and continuum), the idea behind the method is to constrain the sign or phase of the overlap of the sampled Slater determinants with a trial wave function. The constraint eliminates the sign or phase instability and restores low power (typically to the third power of system size) computational scaling. Applications to systems ranging from lattice models~\cite{PhysRevLett.104.116402,Shi2013} of correlated systems to solids~\cite{PhysRevB.80.214116,Ma_Ex_Solids_2012} to atoms and molecules~\cite{Wirawan_CaH2} have shown that these methods are very accurate, even with simple trial wave functions taken directly from Hartree-Fock or density-functional calculations.

Since these methods combine standard mean-field approaches with stochastic QMC, they pose a formidable barrier to beginners. As such, it is useful to have a pedagogical platform to learn the methods and to aid further code development. In this paper, we present a \Matlab{} package that fulfills these roles. The package illustrates the CPMC method for the Hubbard model, with a graphical interface. The ground-state energy is calculated using importance sampling and implementing the full algorithmic details. With this open-source package, calculations can be performed directly on Hubbard-like models in any dimensions, under any boundary conditions. It will be straightforward to generalize the code for applications in many other models of correlated electron systems. Furthermore, the code contains the core QMC algorithmic ingredients  for  a total energy AFQMC calculation. These ingredients can be combined with standard electronic structure machineries~\cite{Martin_book} (see, e.g., Refs~\cite{ABINIT_CPC} and~\cite{NWChem_CPC}) to develop a production code for AFQMC total energy calculations in molecules and solids.

The tool presented here allows users to experiment with various model and run parameters and visualize the results. For this purpose \Matlab{} offers many advantages over traditional programing languages such as \textsc{FORTRAN} or \textsc{C}. As an interpreted language, \Matlab{} requires no compilation, is platform-independent and allows easy interaction with the algorithm during runtime. \Matlab{} also provides an array of tools to visualize results from computations including a full graphical user interface (GUI). These advantages make the package a better choice for our purposes, as discussed above, than regular ``production'' codes, despite a large discrepancy in computational speed. We provide several examples and include more questions in the exercises, which illustrate the usage of the code, key algorithmic features, how to compute various properties in the Hubbard model, and how to generalize it for other applications. As a pedagogical tool, the package can be used in combination with the lecture notes in Ref.~\cite{Zhang2013-Juelich-LectNotes} and references therein.

The remainder of the paper is organized as follows. In Section~\ref{sec: method} we introduce the formalism and various technical ingredients of the CPMC method, and notations used in the rest of this paper. In Section~\ref{sec: algorithm} the actual algorithm is outlined in detail. An overview of the software package and instructions is given in Section~\ref{sec: package}, including a list of exercises. In Section~\ref{sec:speed}, we briefly discuss the computational cost and comment on the relation between this package and one written in a more conventional numerical programming language. Then in Section~\ref{sec:examples}, we present several applications which illustrate algorithmic issues; benchmark the method; discuss CPMC calculations of the kinetic energy, potential energy and double occupancy; and report several new CPMC results on the spin- and charge-gaps. We summarize in Section~\ref{sec:summary} and include a list of some useful \Matlab{} commands in~\ref{sec:matlabCommands}.

\section{Method and Notation}
\label{sec: method}

The ground-state CPMC algorithm has two main components: The \emph{first} component is the formulation of the ground state projection as an open-ended importance-sampled random walk. This random walk takes place in Slater determinant space rather than configuration space like in Green's function Monte Carlo (GFMC)~\cite{GFMC-1,GFMC-2}. There are two ways to perform this random walk. Traditional projector QMC uses an exact unconstrained projection that suffers from exponential scaling in computational cost with increasing system size due to the sign problem. In contrast, CPMC achieves polynomial scaling by using an approximate constrained projection, which becomes exact if the trial wave function (used to impose the constraint) is identical to the ground state. Furthermore, importance sampling makes CPMC a more efficient way to do projector QMC in many cases. The \emph{second} component is the constraint of the paths of the random walk so that any Slater determinant generated during the random walk maintains an appropriate overlap with a known trial wave function $\ket{\psi\ped{T}}$. This constraint eliminates the sign decay, making the CPMC method scale algebraically instead of exponentially, but introduces a systematic error in the algorithm. These two components are independent of each other, and can be used separately. We call the combination of these two components the ground-state CPMC algorithm. This section only briefly reviews the application of CPMC to the Hubbard model. The reader is referred to Refs.~\cite{cpmc0K-2,Zhang2013-Juelich-LectNotes} and references therein for a discussion of the Hubbard model implementation and generalization of this theory to other, more realistic Hamiltonians.
 
\subsection{Slater determinant space}

The CPMC method works with a chosen one-particle basis. A Born-Oppenheimer Hamiltonian with standard electronic interactions does not mix the spins. We assume in the discussion below that the Hamiltonian conserves total $\hat S_z$, and that the number of electrons  with each spin component is fixed. It is straightforward to treat a Hamiltonian that does mix the spin species~\cite{Hao_PHF_2014_preprint}.

We will use the following notation:

\begin{itemize}
\item $M$: the number of single-electron basis states i.e.\ the number of lattice sites in the Hubbard model.
\item $\ket{\chi_i}$: the $i$-th single-particle basis
state ($i=1,2, \dots ,M$). 

\item $c^\dagger_i$ and $c_i$: creation and annihilation operators for an electron in $\ket{\chi_i}$. $n_i\equiv c^\dagger_i c_i$ is the corresponding number operator.

\item $N$: the number of electrons. $N_\sigma$ is the number of electrons with spin $\sigma$ ($\sigma=\uparrow$ or $\downarrow$). As expected, $N_\sigma\le M$.

\item $\varphi$: a single-particle orbital. The coefficients in the expansion
 $
 \varphi=\sum_i \varphi_{i} \ket{\chi_i}
=\sum_i c^\dagger_i \varphi_{i} \ket{0}
$ in the single particle basis $\{ \ket{\chi_i} \}$ can be conveniently expressed as an $M$-dimensional vector: 

\beq
\begin{pmatrix}
\varphi_{1} \\
      \varphi_{2} \\
      \vdots \\
      \varphi_{M}
\end{pmatrix} \label{eqn:singleParticleOrbital}
\eeq

\item $\ket{\phi}$: a many-body wave function which can be written as a Slater determinant. Given $N$ different single-particle orbitals, we form a many-body wave function from their anti-symmetrized product:
\beq
\ket{\phi} \equiv \hat{\varphi}_1^\dagger 
\hat{\varphi}_2^\dagger \cdots
\hat{\varphi}_N^\dagger\ket{0}
\label{eq:slater}
\eeq
where the operator $\hat{\varphi}_m^\dagger \equiv \sum_i c_i^\dagger\, \varphi_{i,m}$ creates an electron in the $m$-th single-particle orbital as described in Eq.~\eqref{eqn:singleParticleOrbital}. 

\item $\Phi$: an $M \times N$ matrix which represents the coefficients of the orbitals used to construct a Slater determinant $\ket{\phi}$:
\beq
\Phi\equiv
\begin{pmatrix}
\varphi_{1,1} & \varphi_{1,2} & \cdots & \varphi_{1,N}\\
\varphi_{2,1} & \varphi_{2,2} & \cdots & \varphi_{2,N}\\
\vdots & \vdots & & \vdots\\
\varphi_{M,1} & \varphi_{M,2} & \cdots & \varphi_{M,N}
\end{pmatrix}
\label{eqn:slaterMatrix}
\eeq
Each column of this matrix is an $M$-dimensional vector and represents a single-particle orbital  described by Eq.~\eqref{eqn:singleParticleOrbital}. For brevity, we will subsequently refer to this $M \times N$ matrix as a Slater determinant.

\item $\ket{\Psi}$ (upper case): a many-body wave function which is not necessarily a single Slater determinant, e.g.\ the many-body ground state $\ket{\Psi_0}$.

\end{itemize}

We list several properties of a Slater determinant. First, for any two non-orthogonal Slater determinants, $\ket{\phi}$ and $\ket{\phi'}$, it can be shown that their overlap integral is given by a number:
\begin{equation}
 \braket{\phi|\phi'} = \det\left(\Phi\ap{\dagger}\Phi'\right),
\label{eq:ovlp}
\end{equation}
where $\Phi\ap{\dagger}$ is the conjugate transpose of the matrix $\Phi$.

Second, the operation on any Slater determinant in Eq.~\eqref{eq:slater} by the exponential of a one body operator
\begin{equation}
\hat{B}=\exp \left(\sum_{ij}^M c_i^\dagger U_{ij}c_j\right)
\label{eq:spo}
\end{equation}
simply leads to another Slater determinant~\cite{PhysRevB.41.11352}:
\begin{equation}
\hat{B}\ket{\phi} =
	\hat{\phi}_1^{\prime\,\dagger} \hat{\phi}_2^{\prime\,\dagger} \cdots
	\hat{\phi}_N^{\prime\;\dagger}\,\ket{0}
	\equiv \ket{\phi'}
\label{eq:expo}
\end{equation}
with $\hat{\phi}_m^{\prime\;\dagger} = \sum_j c_j^\dagger\,\Phi'_{jm}$ and $\Phi' \equiv \e^{U}\Phi$, where the matrix $U$ is formed from elements $U_{ij}$. Since $B\equiv \e^{U}$ is  an $M \times M$ square matrix, the operation of $\hat{B}$ on $\ket{\phi}$ simply involves multiplying $\e^U$, an $M\times M$ matrix, to $\Phi$, an $M\times N$ matrix.

As mentioned above, operations on the spin-up sector  do not affect the spin-down sector and vice versa. Thus it is convenient to represent each Slater determinant as two independent spin parts:
\beq
\ket{\phi} = \ket{\phi^\uparrow}  \otimes \ket{\phi^\downarrow}
\label{eqn:spinDeterminants}
\eeq
The corresponding matrix representation is 
\beq
\Phi=\Phi^\uparrow \otimes \Phi^\downarrow
\label{eqn:spinDetMatrices}
\eeq
where $\Phi^\uparrow$ and $\Phi^\downarrow$ have dimensions $M\times N_\uparrow$ and $M\times N_\downarrow$, respectively. The overlap between any two Slater determinants is simply the product of the overlaps of individual spin determinants:
\beq
\braket{\phi | \phi'} = \prod_{\sigma=\uparrow,\downarrow} \braket{\phi^\sigma | \phi^{\prime\sigma}}
=\det\left[(\Phi^\uparrow)\ap{\dagger}\Phi'^\uparrow\right] \cdot \det\left[(\Phi^\downarrow)\ap{\dagger}\Phi'^\downarrow\right]\,.
\eeq
Any operator $\hat{B}$ described by Eq.~\eqref{eq:spo} acts independently on the two spin parts:
\beq
\hat{B} \ket{\phi} = \hat{B}^\uparrow \ket{\phi^\uparrow} \otimes \hat{B}^\downarrow \ket{\phi^\downarrow} \label{eqn:operatorDiffSpins}
\eeq

Each of the spin components of $\hat{B}$ can be represented as an $M\times M$ matrix. Applying  $\hat{B}$ to a Slater determinant simply involves matrix multiplications for the $\uparrow$ and $\downarrow$ components separately, leading to another Slater determinant $\ket{\phi'}$ as in Eq.~\eqref{eq:expo} i.e.\ the result is $B^\uparrow \Phi^\uparrow \otimes B^\downarrow \Phi^\downarrow$. Unless specified, the spin components of $\hat{B}$ are identical i.e.\ $B^\uparrow = B^\downarrow$ (note the absence of a hat on $B$ to denote the matrix of the operator $\hat B$). 

\subsection{The Hubbard Hamiltonian}

The one-band Hubbard model is a simple paradigm of a system of interacting electrons. Its Hamiltonian is given by
\begin{equation}
  \hat H = \hat K+ \hat V =-t\sum_{\langle ij \rangle \sigma} (c_{i \sigma}^\dagger c_{j\sigma}^{\phantom{\dagger}} + c_{j \sigma}^\dagger c_{i\sigma}^{\phantom{\dagger}} ) + U \sum_i n_{i \uparrow} n_{i \downarrow},
\label{eq:H}
\end{equation}
where $t$ is the hopping matrix element, and $c_{i \sigma}^\dagger$ and $c_{i \sigma}$ are electron creation and destruction operators, respectively, of spin $\sigma$ on site $i$. The Hamiltonian is defined on a lattice of dimension $M=\prod_d L_d$. The lattice sites serve as the basis functions here, i.e., $\ket{\chi_i}$ denotes an electron localized on  the site labeled by $i$. The notation $\langle\;\rangle$ in Eq.~\eqref{eq:H} indicates nearest-neighbors.  The on-site Coulomb repulsion is $U>0$, and the model only has two parameters: the strength of the interaction $U/t$ and the electron density $(N_\uparrow + N_\downarrow)/M$. In this paper we will use $t$ as the unit of energy and set $t=1$. 

The difference between the Hubbard Hamiltonian and a general electronic Hamiltonian is in the structure of the matrix elements in $\hat K$ and $\hat V$. In the latter, $\hat K$ is specified by hopping integrals of the form $K_{ij}$, while $\hat V$ is specified by Coulomb matrix elements of the form $V_{ijkl}$, with $i, j, k, l$ in general running from $1$ to $M$. In terms of the CPMC method, the structure of $\hat K$ makes essentially no difference. The structure of $\hat V$, however, dictates the form of the Hubbard-Stratonovich transformation (see Section~\ref{sec:hubbardStratonovichTransformation}) . For the Hubbard interaction, the resulting one-body propagators turn out to be real, as shown below, while for the general case complex propagators arise and cause a phase problem~\cite{Zhang2013-Juelich-LectNotes}.

\subsection{Ground-state projection}

We will focus on ground-state calculations in this paper. (Finite-temperature generalizations to the grand-canonical ensemble also exist~\cite{Zhang1999_FTCPMC,Rubenstein2012}.)
The ground-state wave function $\ket{\Psi_0}$ can be obtained from any trial wave function $\ket{\Psi\ped{T}}$ that is not orthogonal to $\ket{\Psi_0}$ by repeated applications of the ground-state projection operator 
\beq
\mathcal{P}\ped{gs}=\e^{-\Delta\tau (\hat H-E\ped{T})} \label{eqn:projectionOperator}
\eeq
where $E\ped{T}$ is the best guess of the ground-state energy. That is, if the wave function at the $n$-th time step is $\ket{\Psi^{(n)}}$, the wave function at the next time step is given by
\beq
\ket{\Psi^{(n+1)}} = \e^{-\Delta\tau (\hat H-E\ped{T})} \ket{\Psi^{(n)}} \label{eq:process}
\eeq

With a small $\Delta\tau$, the second-order Trotter approximation can be used:
\beq
\e^{-\Delta\tau \hat H}=\e^{-\Delta\tau(\hat K+\hat V)} \approx \e^{-\Delta\tau\, \hat K/2} \e^{-\Delta\tau\, \hat V} \e^{-\Delta\tau\, \hat K/2}\,. \label{eqn:trotterBreakup}
\eeq
The residual Trotter error can be removed by, for example, extrapolation with several independent runs of sufficiently small $\Delta \tau$ values. We illustrate this technique in an exercise in Section~\ref{sec:timeStepExtrapolation}.

\subsubsection{The Hubbard-Stratonovich transformation} \label{sec:hubbardStratonovichTransformation}

In Eq.~\eqref{eqn:trotterBreakup}, the kinetic energy (or, more generally, one-body) propagator $\hat B\ped{K/2}\equiv \e^{-\Delta\tau \hat K/2}$ has the same form as Eq.~\eqref{eq:spo}. However, the potential energy propagator $\e^{-\Delta\tau \hat V}$ does not. A Hubbard-Stratonovich (HS) transformation can be employed to transform $\e^{-\Delta\tau \hat V}$ into the desired form. In the Hubbard model, we can use the following:

\begin{equation}
e^{-\Delta\tau U n_{i \uparrow} n_{i \downarrow}} =  \e^{-\Delta\tau U (n_{i\uparrow}+n_{i \downarrow})/2} \sum_{x_i=\pm 1}  p(x_i) \, \e^{\gamma x_i (n_{i \uparrow}-n_{i \downarrow})}\,,
\label{eq:HSdiscrete}
\end{equation}
where $\gamma$ is given by $\cosh(\gamma) = \exp (\Delta\tau U/2)$. We interpret $p(x_i)=1/2$ as a discrete probability density function (PDF) with $x_i=\pm 1$. 

In Eq.~\eqref{eq:HSdiscrete}, the exponent on the left, which comes from the interaction term $\hat V$ on the $i$-th site, is quadratic in $n$, indicating the interaction of \emph{two} electrons. The exponents on the right, on the other hand, are linear in $n$, indicating two non-interacting electrons in a common external  field characterized by $x_i$. Thus an interacting system has been converted into a \emph{non-interacting} system living in fluctuating external \emph{auxiliary fields} $x_i$, and the summation over all such auxiliary-field configurations recovers the many-body interactions. The special form of HS for the Hubbard interaction  is due to Hirsch~\cite{Hirsch}. The linearized operator on the right hand side in Eq.~\eqref{eq:HSdiscrete} is the spin  $(n_{i \uparrow}-n_{i \downarrow})$ on each site.

In this paper and in the code, we will use the discrete spin decomposition in Eq.~\eqref{eq:HSdiscrete}. There exist other ways to do the HS transformation, e.g.\ based on the Gaussian integral:
\beq
\e^{\hat{A}^2} =  \frac{1}{\sqrt{2\pi}} \int_{-\infty}^\infty \du x \, \e^{-x^2/2} \; \e^{\sqrt{2}x\, \hat{A}}\,.
\eeq
There is also a charge version of the discrete HS transformation involving the linearized operator 
$n_{i\uparrow}+n_{i\downarrow}$, the total charge on each site.

Different forms of the HS transformation can have different efficiencies in different situations.  In 
particular, preserving the appropriate symmetries of the system can significantly reduce the statistical fluctuations and reduce the error from the constrained-path approximation~\cite{Shi2013}.

Since we represent a Slater determinant as individual spin determinants in Eq.~\eqref{eqn:spinDeterminants}, it is convenient to spin-factorize Eq.~\eqref{eq:HSdiscrete} as
\beq
\e^{-\Delta\tau U n_{i \uparrow} n_{i \downarrow}} =\sum_{x_i=\pm 1} p(x_i) \left[\hat b^\uparrow_V (x_i) \otimes \hat b^\downarrow_V (x_i) \right]
\label{eqn:HSDiscreteSpins}
\eeq
where the spin-dependent operator $\hat b^\sigma_V (x_i)$ on the $i$-th lattice site is defined as
\beq
\hat b^\sigma\ped{V} (x_i) = \e^{-[\Delta\tau U/2 - s(\sigma)\, \gamma \, x_i] c_{i\sigma}^\dagger c_{i\sigma}}
\eeq
and $s(\uparrow)=+1$ and $s(\downarrow)=-1$. The related operator $\hat b\ped{V} (x_i)$ (i.e.\ without $\sigma$) includes both the spin up and spin down parts. Below we will  use the corresponding symbol without hat ($b\ped{V}$) to denote the matrix representation of the operator ($\hat b\ped{V}$) associated with that symbol.

The potential energy propagator $\e^{-\Delta\tau \hat V}$ over all sites can easily be seen to be the product of the propagators $\e^{-\Delta\tau U n_{i\uparrow} n_{i\downarrow}}$ over each site:
\beq
\e^{-\Delta\tau \hat V} 
=\sum_{\vv{x}} P(\vv{x}) \prod_{\sigma=\uparrow,\downarrow} \hat B^\sigma_V(\vv{x})
\eeq
where $\vv{x}=\{x_1,\dots,x_M \}$ is one configuration  of  auxiliary fields over all $M$ sites and $\hat B^\sigma_V(\vv{x})=\prod_i \hat b^\sigma_V (x_i)$ is the $\vv{x}$-dependent product of the spin-$\sigma$ propagators over all sites. The overall PDF here is $P(\vv{x})=\prod_i p(x_i)=\left( \frac{1}{2}\right)^M$, to be distinguished from the PDF $p$ for one individual auxiliary field $x_i$  in Eq.~\eqref{eqn:HSDiscreteSpins}.

Now the projection operator in Eq.~\eqref{eqn:projectionOperator} can be expressed entirely in terms  of operators in Eq.~\eqref{eq:spo}
\beq
\mathcal{P}\ped{gs} \approx \e^{\Delta\tau E\ped{T}} \sum_{\vv{x}} P(\vv{x}) \prod_{\sigma=\uparrow,\downarrow}
\hat B^\sigma\ped{K/2} \hat B^\sigma\ped{V}(\vv{x}) \hat B^\sigma\ped{K/2} 
\label{eqn:propagatorWithHS}
\eeq
As noted in Eq.~\eqref{eqn:operatorDiffSpins}, $B\ped{K/2} $ has an  $\uparrow$ and a $\downarrow$ component, each of which is an $M\times M$ matrix. Applying each $\hat B\ped{K/2} $ to a Slater determinant $\ket{\phi}$ simply involves matrix multiplications with the matrix $B\ped{K/2} $ for the $\uparrow$ and $\downarrow$ components of $\Phi$ separately, leading to another Slater determinant $\ket{\phi'}$ as in Eq.~\eqref{eq:expo}.

\subsubsection{A toy model for illustration} \label{sec:toyModel}

Let us take, for example, a simple one-dimensional four-site Hubbard model with $N_\uparrow=2$, $N_\downarrow=1$ and  open boundary condition. The sites are numbered sequentially. 

First let us examine the trivial case of free electrons i.e.\ $U=0$. We can write down the \emph{one-electron} Hamiltonian matrix, which is of dimension $4\times 4$:
\beq
H = 
\begin{pmatrix}
0 & -1 & 0 & 0\\
-1 & 0 & -1 & 0\\
0 & -1 & 0 & -1\\
0 & 0 & -1 & 0
\end{pmatrix}
\eeq
Direct diagonalization gives us the eigenstates of $H$ from which we immediately obtain the matrix $\Phi_0$ for the ground-state wave function $\ket{\psi_0}$: 
\beq
\Phi_0 = 
\begin{pmatrix}
0.3717 & -0.6015\\
0.6015 & -0.3717\\
0.6015 & \phantom{-}0.3717\\
0.3717 & \phantom{-}0.6015
\end{pmatrix}
\otimes
\begin{pmatrix}
0.3717\\
0.6015\\
0.6015\\
0.3717
\end{pmatrix} \label{eqn:egSlaterMatrix}
\eeq
where the first matrix contains two single-particle orbitals (two columns) for the two $\uparrow$ electrons and the second matrix contains one single-electron orbital for the one $\downarrow$ electron. Each single-electron orbital is an eigenvector of $H$. 

The matrix $\Phi_0$ represents $\ket{\phi_0}$ in the same way that Eq.~\eqref{eqn:slaterMatrix} represents Eq.~\eqref{eq:slater}. Of course, the solution above is also the restricted Hartree-Fock solution to the \emph{interacting} problem. We will often use $\Phi_0$ as the trial wave function in CPMC below.

Next we consider the interacting problem, with $U>0$. Applying the HS transformation of Eq.~\eqref{eq:HSdiscrete} to Eq.~\eqref{eqn:trotterBreakup}, we have

\begin{widetext}
\begin{equation}
\mathcal{P}\ped{gs} = \e^{\Delta\tau E\ped{T}-\Delta\tau U (N_{\uparrow}+N_{\downarrow})/2}  
\sum_{\vec x} P(\vec x) \; 
 B\ped{K/2} \cdot
\begin{pmatrix}
\e^{\gamma x_1} & 0 & 0 & 0 \\
0 & \e^{\gamma x_2} & 0 & 0 \\
0 & 0 & \e^{\gamma x_3} & 0 \\
0 & 0 & 0 & \e^{\gamma x_4}
\end{pmatrix} 
\cdot B\ped{K/2} \nonumber \\
 \otimes 
B\ped{K/2} \cdot
\begin{pmatrix}
\e^{-\gamma x_1} & 0 & 0 & 0 \\
0 & \e^{-\gamma x_2} & 0 & 0 \\
0 & 0 & \e^{-\gamma x_3} & 0 \\
0 & 0 & 0 & \e^{-\gamma x_4}
\end{pmatrix}
\cdot B\ped{K/2}\,, 
\tag{23} 
\label{eqn:HSFourSite}
\end{equation}
\end{widetext}
\setcounter{equation}{23} 

where $\vec x= \{x_1,x_2,x_3,x_4\}$ and $P(\vv{x})=\left( \frac{1}{2}\right)^4$. This is just Eq.~\eqref{eqn:propagatorWithHS} specialized to a four-site lattice.

\subsection{Random walk in Slater determinant space}\label{sec:randomWalk}

The first component of the CPMC algorithm is the reformulation of the projection process as branching, open-ended random walks in Slater determinant space (instead of updating a fixed-length path in auxiliary-field space).

Let us define $B\ped{V}(\vv{x})=\prod_\sigma B^\sigma_V(\vv{x})$ as we have done for $\hat b\ped{V}$. Applying the HS-transformed propagator in Eq.~\eqref{eqn:propagatorWithHS}  to one projection step in Eq.~\eqref{eq:process} gives
\beq
\ket{\phi^{(n+1)}} =  \e^{\Delta\tau E\ped{T}} \sum_{\vv{x}} P(\vv{x})   \left[ \hat B\ped{K/2} \hat B\ped{V}(\vv{x}) \hat B\ped{K/2} \right] \ket{\phi^{(n)}}\,. \label{eq:cpmc}
\eeq

In the Monte Carlo (MC) realization of this iteration, we represent the wave function at each stage by a finite ensemble of Slater determinants, i.e.
\begin{equation}
\ket{\Psi^{(n)} } \propto \sum_k \ket{ \phi^{(n)}_k }
\label{eq:wf}
\end{equation}
where $k$ labels the Slater determinants and an overall normalization factor of the wave function has been omitted. These Slater determinants will be referred to as \emph{random walkers}. 

The iteration in Eq.~\eqref{eq:cpmc} is achieved stochastically by MC sampling of $\vv{x}$. That is, for each random walker $\ket{\phi^{(n)}_k}$, we choose an auxiliary-field configuration $\vv{x}$ according to the PDF $P(\vv{x})$ and  propagate the determinant to a new determinant via $\ket{\phi^{(n+1)}_k}=\hat B\ped{K/2} \hat B\ped{V}(\vv{x}) \hat B\ped{K/2}\ket{\phi^{(n)}_k}$. 

We repeat this procedure for \emph{all} walkers in the population. These operations accomplish one step of the random walk. The new population represents $\ket{\Psi^{(n+1)}}$ in the sense of Eq.~\eqref{eq:wf}, i.e.\ $\ket{\Psi^{(n+1)}} \propto \sum_k \ket{\phi_k^{(n+1)}}$. These steps are iterated until sufficient data has been collected. After an equilibration phase, all walkers thereon are MC samples of the ground-state wave function $\ket{\Psi_0}$ and ground-state properties can be computed. We will refer to this type of approach as \emph{free projection}. In practice, branching occurs because of the re-orthonormalization of the walkers, which we discuss below in Section~\ref{subset:implement-issues}. We emphasize that the statistical error bar can be reduced significantly with more ``optimal'' forms of HS transformations to extend the reach of free-projection calculations  (see e.g., Refs.~\cite{Zhang2013-Juelich-LectNotes,Shi2013}).

\subsection{Importance sampling}

To improve the efficiency of Eq.~\eqref{eq:cpmc} and make it a practical and scalable algorithm, an importance sampling scheme~\cite{Zhang2013-Juelich-LectNotes,GFMC-1,GFMC-2} is required. In the procedure just described above, no information is contained in the sampling of $\vv{x}$ on the importance of the resulting determinant in representing $\ket{\Psi_0}$. Computing the mixed estimator of the ground-state energy 
\beq
E\ped{mixed}\equiv \frac{\braket{ \phi\ped{T} |\hat H|\Psi_0}}{\braket{ \phi\ped{T} |\Psi_0}}
\label{eq:mixed-est}
\eeq
requires estimating the denominator by $\sum_k \braket{ \phi\ped{T} |\phi_k}$ where $\ket{\phi_k}$ are random walkers after equilibration. Since these walkers are sampled with no knowledge of $\braket{\phi\ped{T}|\phi_k}$, terms in the summation over $\ket{\phi_k}$ can have large fluctuations that lead to large statistical errors in the MC estimate of the denominator, thereby in that of $E\ped{mixed}$.

With importance sampling,  first we define an importance function:
\begin{equation}
O\ped{T}(\phi_k)\equiv \braket{ \phi\ped{T}|\phi_k}\,,
\label{eq:O_T}
\end{equation}
which estimates the overlap of a Slater determinant $\ket{\phi}$ with the ground-state wave function (approximated here by a trial wave function). As in Diffusion Monte Carlo (DMC)~\cite{Umrigar1993}, we also assign a weight $w_k=O\ped{T}(\phi_k)$ to each walker. This weight is initialized to unity for all walkers since  in the initial ensemble, $\ket{\phi_k^{(0)}}=\ket{\phi\ped{T}}$ for all $k$. 

We then iterate a \emph{formally} different but \emph{mathematically} equivalent  version of Eq.~\eqref{eq:cpmc}: 

\beq
 \ket{\widetilde{\phi}^{(n+1)}} \gets \sum_{\vv{x}} 
\widetilde{P}(\vv{x}) \hat B(\vv{x})  \ket{\widetilde{\phi}^{(n)}}
 \label{eq:impcpmc}
\eeq
where
\beq
\hat B(\vv{x}) 
= \hat B\ped{K/2} \hat B\ped{V}(\vv{x}) \hat B\ped{K/2} \,.
\eeq
The walkers $\ket{\widetilde{\phi}^{(n)}}$ are now sampled from a new distribution. They schematically represent the ground-state wave function by: 
\begin{equation}
\ket{\Psi^{(n)} } \propto \sum_k  w^{(n)} \frac{\ket{\phi^{(n)}_k}}{O\ped{T}(\phi^{(n)}_k)}\,, 
\label{eqn:wf-imp}
\end{equation}
in comparison to Eq.~\eqref{eq:wf}.

The modified function $\widetilde{P}(\vv{x}) $ in Eq.~\eqref{eq:impcpmc} is
$\widetilde{P}(\vv{x}) = \prod_i^M  \widetilde{p}(x_i)$, where  the probability for sampling the auxiliary-field at each lattice site is given by
\begin{equation}
\widetilde{p}(x_i)=\frac{O\ped{T}(\phi^{(n)}_{k,i})}{O\ped{T}(\phi^{(n)}_{k,i-1})} \, p(x_i).
\label{eq:impcpmc_p}
\end{equation}
where 
\begin{equation}
\ket{\phi^{(n)}_{k,i-1}}=\hat b_V(x_{i-1})\,\hat b_V(x_{i-2})\,\cdots\,\hat b_V(x_{1})\,\ket{\phi^{(n)}_{k}} 
\label{eq:impcpmc_substep}
\end{equation}
is the current
state of the $k$-th walker,  ${\ket{\phi^{(n)}_k}}$, after its first $(i-1)$ fields have been sampled and updated, and  
\begin{equation}
\ket{\phi^{(n)}_{k,i}}=\hat b_V(x_i)\ket{\phi^{(n)}_{k,i-1}}
\label{eq:impcpmc_substep2}
\end{equation}
is the next sub-step after the $i$-th field is selected and the walker is updated. Note that in the notation above $\ket{\phi^{(n)}_{k,0}}=\ket{\phi^{(n)}_{k}}$ and
$\ket{\phi^{(n)}_{k,M}}=\ket{\phi^{(n+1)}_{k}}$. As expected, $\widetilde{P}(\vv{x})$ is a function of both the current and future positions in Slater-determinant space. Further, $\widetilde{P}(\vv{x})$ modifies $P(\vv{x})$ such that the probability of sampling $\vv{x}$ is increased when $\vv{x}$ leads to a determinant with larger overlap with $\ket{\phi\ped{T}}$ and is decreased otherwise. In each $\widetilde{p}(x_i)$, $x_i$ can only take the value of $+1$ or $-1$ and can be sampled by a heatbath-like algorithm: choosing $x_i$ from the PDF $\widetilde{p}(x_i)/N_i$ where the normalization factor is
\beq
\mathcal{N}_i \equiv \widetilde{p}(x_i=+1)+\widetilde{p}(x_i=-1)\,,
\eeq
and carrying a weight for the walker 
\beq
w^{(n)}_{k,i}=\mathcal{N}_i w^{(n)}_{k,i-1}\,, 
\eeq
in which we have used the same notation as in Eqs.~\eqref{eq:impcpmc_substep} and \eqref{eq:impcpmc_substep2}. The ratio of the overlaps in Eq.~\eqref{eq:impcpmc_p}, involving a change of only one site (or even a few sites if desired), can be computed quickly using the Sherman-Morrison formula, as shown in the code. The inverse of the overlap matrix $\left[(\Phi\ped{T})^\dagger  \Phi_{k}^{(n)}\right]^{-1}$ is kept and updated after each $x_i$ is selected.

It should be pointed out that a significant  reduction  in computational cost is possible in the present code. Because the free-electron trial wave function is used here, the overlap matrix can be trivially updated after each application of $\hat B\ped{K/2}$. In the code, however, the update is still carried out explicitly in order to allow for a more general form of the trial wave function.

We note that, for a general continuous auxiliary-field, the importance sampling can be achieved by a force bias~\cite{PhysRevLett.90.136401,PhysRevE.70.056702,Zhang2013-Juelich-LectNotes}. The above discrete version can be viewed as a two-point realization of the continuous case.

\subsection{The sign problem and the constrained path approximation}\label{sec:constrainedPath}

\subsubsection{The sign problem}
The sign problem occurs because of the fundamental symmetry between the fermion ground state $\ket{\Psi_0}$ and its negative $-\ket{\Psi_0}$~\cite{inter-1}. This symmetry implies that, for any ensemble of Slater determinants $\{\ket{\phi}\}$ which gives a Monte Carlo representation of the ground-state wave function, there exists another ensemble $\{-\ket{\phi}\}$ which is also a correct representation.
In other words, the Slater determinant space can be divided into two degenerate halves ($+$ and $-$) whose bounding surface $\mathcal{N}$ is defined by $\braket{\Psi_0|\phi}=0$. This surface  is in general \emph{unknown}. Except for some special cases~\cite{Hirsch}, walkers do cross $\mathcal{N}$ in their propagation by $\mathcal{P}\ped{gs}$, causing the sign problem. At the instant such a walker lands on $\mathcal{N}$, the walker will make no further contribution to the representation of the ground state at any later time because
\begin{equation}
\braket{ \Psi_0|\phi } = 0 \ \implies \braket{\Psi_0|\e^{-\tau H}|\phi}= 0 \quad\text{for any } \tau
\label{eq:node}
\end{equation}

Paths that result from such a walker have equal probabilities of being in either half of the Slater determinant space \cite{Zhang2013-Juelich-LectNotes}. Computed analytically, they would cancel and make no contribution in the ground-state wave function. However, because  the random walk has no knowledge of $\mathcal{N}$, these paths continue to be sampled (randomly) in the random walk and become Monte Carlo noise.

To eliminate the decay of the signal-to-noise ratio, we impose the constrained path approximation. It requires that each random walker at each step have a positive overlap with the trial wave function $\ket{\phi\ped{T}}$:
\begin{equation}
\braket{ \phi\ped{T}|\phi_k^{(n)} } > 0.
\label{eq:constraint}
\end{equation}
This yields an approximate solution to the ground-state wave function, $\ket{\Psi\ap{c}_0 } =\sum_\phi \ket{\phi }$, in which all Slater determinants $\ket{\phi}$ satisfy Eq.~\eqref{eq:constraint}. Note that from Eq.~\eqref{eq:node}, the constrained path approximation becomes \emph{exact} for an exact trial wave function $\ket{\psi\ped{T} }=\ket{\Psi_0 }$. The constrained path approximation is  easily implemented by redefining the importance function in Eq.~\eqref{eq:O_T}:
\begin{equation}
O\ped{T}(\phi_k)\equiv \operatorname{max}\{ \braket{ \phi\ped{T}| \phi_k }, 0\}
\label{eq:imp_cp}
\end{equation}
This prevents walkers from crossing the trial nodal surface ${\cal N}$ and entering the ``$-$'' half-space as defined by $\ket{\phi\ped{T}}$. We note that imposing Eq.~\eqref{eq:constraint} is fundamentally different from just discarding the negative contributions in the denominator of Eq.~\eqref{eq:mixed-est}; the constrained path condition results in a distribution of walkers that  \emph{vanishes smoothly} at the interface between the ``$+$'' and ``$-$'' parts of the determinant space. The use of a finite $\Delta\tau$ causes a small discontinuity which is a form of Trotter error in the constraint; this error can be further reduced~\cite{Zhang_CPC_FT}.

\subsubsection{Twist boundary condition and the phase problem} 

Our simulations are carried out in supercells of finite sizes. For most 
quantities that we wish to calculate, the periodic boundary condition (PBC) causes large finite-size effects,  often compounded by significant shell effects~\cite{PhysRevB.78.165101}. In order to reduce these effects and reach the thermodynamic limit more rapidly, it is more effective to use the twist boundary condition (TBC) and average over the twist angles~\cite{PhysRevE.64.016702,PhysRevB.78.165101} (TABC). Under the TBC, the wave function $\Psi (\mathbf{r_1},\mathbf{r_2},\dots,\mathbf{r_N})$ gains a phase when electrons hop around lattice boundaries:
\beq
\Psi(\dots , \mathbf{r_j+L},\dots ) = \e^{\iu \, \mathbf{\widehat{L}} \cdot \bm{\Theta} } \Psi(\dots , \mathbf{r_j},\dots )\,,
\eeq
where $\mathbf{\widehat{L}}$ is the unit vector along $\mathbf{L}$ and the twist angle $\bm{\Theta}=(\theta_x,\theta_y,\theta_z,\dots)$ is a parameter with $\theta_d\in (-\pi,\pi]$ for $d=x,y,z,\dots$. It is implemented straightforwardly as a modification to the matrix elements in the $\hat K$ part of the Hamiltonian. A twist is equivalent to shifting the underlying momentum space grid by $(\theta_x/L_x,\theta_y/L_y,\theta_z/L_z,\cdots)$. Symmetry can be used to reduce the range of $\bm{\Theta}$. One could either choose to have a special grid of  $\bm{\Theta}$ values~\cite{Martin_book} or choose them randomly~\cite{PhysRevE.64.016702,PhysRevB.78.165101}. In the illustrations we will use the latter. The QMC results are averaged, and the MC error bar will be the combined statistical errors from the random $\bm{\Theta}$ distribution and from each QMC calculation for a particular $\bm{\Theta}$.

With a general twist angle, the AFQMC method will have a ``phase problem'' instead of the sign problem described above. This is because the hopping matrix elements in $\hat K$ now have complex numbers which make the orbitals in the random walkers complex. The stochastic nature of the random walk will then lead to an asymptotic distribution which is symmetric in the complex plane~\cite{PhysRevLett.90.136401,Zhang2013-Juelich-LectNotes}. For each walker $\ket{\phi}$, instead of a $+\ket{\phi}$ and $-\ket{\phi}$ as in the sign problem, there is now an infinite set $\{\e^{\iu \theta}\ket{\phi}\}$ ($\theta \in [0,2\pi]$) from which the random walk cannot distinguish.

The phase problem that occurs   in the present case  is a ``milder'' form of the most general phase problem, because here the phases arise from the one-body hopping term instead of the two-body interaction term. The latter takes place in a long-range Coulomb interaction, for example. In that case, the phase problem is controlled with the phaseless or phase-free approximation~\cite{PhysRevLett.90.136401,Zhang2013-Juelich-LectNotes}. When the phase only enters as a one-body boundary condition, the stochastic auxiliary-fields are not directly coupled to complex numbers. A simple generalization of the constrained-path approximation suffices~\cite{PhysRevB.78.165101}: at each step of propagation, the random walkers are required to satisfy the constraint
\beq
\operatorname{Re}\left\{ \frac{\braket{\phi\ped{T} | \phi_k^{(n+1)}}}{\braket{\phi\ped{T} | \phi_k^{(n)}}} \right\} >0 \label{eqn:phaseConstraint}
\eeq
where $\ket{\phi^{(n)}_k}$ and $\ket{\phi^{(n+1)}_k}$ are the current and proposed walkers. Note that Eq.~\eqref{eq:imp_cp} is a special case of Eq.~\eqref{eqn:phaseConstraint} because $\braket{ \phi\ped{T}|\phi_k^{(0)} }>0$ and all overlaps are real when $\hat K$ is real. We emphasize that, when the HS transformation leads to complex one-body propagators as in the case of realistic electronic problems, an extra step is required to ``importance-transform'' the propagators using the phase of the overlap~\cite{PhysRevLett.90.136401,Zhang2013-Juelich-LectNotes}.

\subsubsection{Systematic error from the constrained-path approximation, and its reduction and removal} \label{sec:phaseProblem}

Most applications have used a single-determinant $\ket{\psi\ped{T}}$ taken directly from a Hartree-Fock (HF) or density-functional theory (DFT) calculations. In the Hubbard model, the restricted HF wave function is the same as the free-electron wave function, while the unrestricted HF wave function breaks spin-symmetry and allows, for example, antiferromagnetism and spin-density-wave states~\cite{HubHF_JXu_0953-8984-23-50-505601}. A large number of benchmarks have been carried out with these wave functions~\cite{Shi2013,Zhang2013-Juelich-LectNotes}.

\begin{figure}
\includegraphics[scale=1]{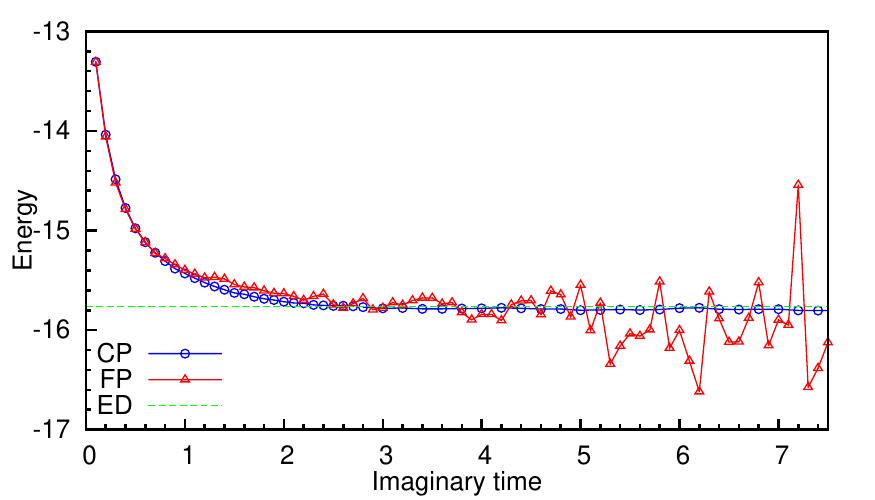}
\caption{(Color online) Illustration of the sign problem. The blue and red curves show the energy during a projection from $\tau=0$ to $\tau=7.5$ with and without the constraint, respectively. The dashed black line shows the exact result. The CP error bars are too small to be seen. To improve clarity, only every other CP energy measurements are shown for $\tau>2.6$.}
\label{fig:signProblem}
\end{figure}

Figure~\ref{fig:signProblem} illustrates the effectiveness of the constrained path approximation. The system is a $4\times 4$ lattice with 7 spin-$\uparrow$ and 7 spin-$\downarrow$ electrons, $U=4$ and $\bm{\Theta}=(0.02\pi,0.04\pi)$. The free-projection (FP) run exhibits growing statistical fluctuations as a function of projection time,  indicative of the sign problem. With the constraint, CPMC fluctuations are always smaller with the same amount of computational cost, and they are independent of projection time. The CPMC results converge to a value below the exact result (horizontal line). 

Multi-determinant trial wave functions can reduce the systematic error because they are better variational wave functions~\cite{2009-C2-exc-afqmc}. Using wave functions that restore symmetries of the system can reduce the systematic error significantly~\cite{Shi2013}. The symmetry restoration can be either in the form of a multi-determinant trial wave function 
or from symmetry projection~\cite{Hao_PHF_2014_preprint}. For example, using a 10-determinant symmetry trial wave function can reduce the systematic error in the CPMC result for the system in Fig.~\ref{fig:signProblem} with periodic	 boundary condition by a significant factor~\cite{Shi2013}.

Recently, it has been demonstrated~\cite{PhysRevB.80.214116,Shi2013,CP_release_Trotter_Sorella_PhysRevB.84.241110} that free-projection and release-constraint calculations allow systematic removal of the constrained-path bias by ``lifting'' the constraint (and bringing back the sign problem). 
This offers another avenue for systematically improvable AFQMC calculations.

\subsection{Energy measurement}
\code{CPMC-Lab} uses the mixed estimator in Eq.~(\ref{eq:mixed-est}) for the ground-state energy which, for an ensemble $\{\ket{\phi} \}$,  is given by:
\beq
E\ped{mixed}
=\frac{\sum_k w_k \, E\ped{L}\left[\phi\ped{T},\phi_k\right]}{\sum_k w_k}
	\label{eqn:mixed-estimator-expanded}
\eeq
where the local energy $E\ped{L}$ is:
\beq
E\ped{L}\left[\phi\ped{T},\phi \right] =\frac{\braket{\phi\ped{T} |\hat H |\phi}}{\braket{\phi\ped{T}  |\phi}}
\eeq
This quantity can be easily evaluated for any walker $\phi$ as follows. For any pair of Slater determinants $\ket{\phi\ped{T}}$ and $\ket{\phi}$, we can calculate the one-body equal-time Green's function as: 
\beq
\braket{c_{j\sigma}^\dagger c_{i\sigma}}\equiv
\frac{\braket{\phi\ped{T} | c_{j\sigma}^\dagger c_{i\sigma} |\phi}}{\braket{\phi\ped{T}  |\phi}}
=\left[ \Phi^\sigma \, [\, (\Phi\ped{T}^\sigma)^\dagger \Phi^\sigma \, ]^{-1} (\Phi\ped{T}^\sigma)^\dagger\right]_{ij}\,.
\label{eqn:greenFunction}
\eeq
This immediately enables the computation of the  kinetic energy term 
$\Braket{ \phi\ped{T} | -t\sum_{\braket{ij}\sigma} c_{i\sigma}^\dagger c_{j\sigma} | \phi }$. 
The potential energy term 
$\Braket{ \phi\ped{T} | U \sum_{i} c_{i\uparrow}^\dagger c_{i\uparrow}^{\phantom{\dagger}} c_{i\downarrow}^\dagger c_{i\downarrow}^{\phantom{\dagger}} | \phi }$ does not have the form of Eq.~\eqref{eqn:greenFunction}, but can be reduced to that form by an application of Wick's theorem:
\begin{align}
\braket{ c_{i\uparrow}^\dagger c_{i\uparrow}^{\phantom{\dagger}} c_{i\downarrow}^\dagger c_{i\downarrow}^{\phantom{\dagger}} } &= 
\braket{c_{i\uparrow}^\dagger c_{i\uparrow}} \braket{c_{i\downarrow}^\dagger c_{i\downarrow}} + 
\braket{c_{i\uparrow}^\dagger c_{i\downarrow}}\braket{c_{i\uparrow}c_{i\downarrow}^\dagger} \nonumber \\
&=\braket{c_{i\uparrow}^\dagger c_{i\uparrow}} \braket{c_{i\downarrow}^\dagger c_{i\downarrow}} 
\end{align}
The reduction to the last line occurs because the $\uparrow$ and $\downarrow$ spin sectors are decoupled in both $\ket{\phi\ped{T}}$ and $\ket{\phi}$. (This is not the case 
in a pairing~\cite{pairing-PsiT} or generalized Hartree-Fock wave function~\cite{Hao_PHF_2014_preprint}. The former is the desired form for $U< 0$. The latter can be used to improve the quality of the trial wave function for $U>0$, and is necessary if the Hamiltonian contains spin-orbit coupling.)

The mixed estimator for the energy arises naturally from importance sampling, and reduces the statistical variance of the computed result. A drawback of the mixed estimator is that the ground-state energy obtained in AFQMC under the constrained path approximation is not variational \cite{Zhang2013-Juelich-LectNotes}. The mixed estimators for observables which do not commute with the Hamiltonian are biased. The back-propagation technique can be used to obtain pure estimates \cite{cpmc0K-2,PhysRevE.70.056702}.

\subsection{Other implementation issues}
\label{subset:implement-issues}

\subsubsection{Population control} \label{sec:popControl}
As the random walk proceeds, some walkers may accumulate very large weights while some will have very small weights. These different weights cause a loss of sampling efficiency because the algorithm will spend a disproportionate amount of time keeping track of walkers that contribute  little to the energy estimate. To eliminate the inefficiency of carrying these weights, a branching scheme is introduced to ``redistribute'' the weights without changing the statistical distribution. In such a scheme, walkers with large weights are replicated and walkers with small weights are eliminated with some probability. 

However, because  branching  might cause the population to fluctuate in an unbounded way (e.g. to grow to infinity or to perish altogether), we perform population control to eliminate this instability at the cost of incurring a bias when the total weight of the walkers is modified. This bias can be reduced by carrying a history of overall weight correction factors. However, the longer this history is included in the energy estimators, the higher the statistical noise. In this package we use a simple ``combing'' method~\cite{PhysRevB.57.11446}, which  discards all history of overall weight normalizations. We note that there exist more elaborate approaches~\cite{Umrigar1993,cpmc0K-2,PhysRevB.57.11446}, for example, keeping a short history of the overall weight renormalization. The length of the history to keep should be a compromise between reducing bias (long) and keeping statistical fluctuation from becoming much larger (short). The effect of population control and how to extrapolate away the bias are illustrated in the Exercises.

\subsubsection{Re-orthonormalization} \label{sec:reorthonormalization}
Repeated multiplications of $B\ped{K/2}$ and $B\ped{V}$ to a Slater determinant in Eq.~\eqref{eq:cpmc} lead to numerical instability, such that round-off errors dominate and $\ket{\phi^{(n)}_k}$ represents an unfaithful propagation of $\ket{\phi^{(0)}_k}$. This instability is controlled by periodically applying the modified Gram-Schmidt orthonormalization to each Slater determinant. For each walker $\ket{\phi}$, we factor its corresponding matrix as $\Phi=QR$ where $R$ is a upper triangular matrix and $Q$ is a matrix whose columns are orthonormal vectors representing the re-orthonormalized single-particle orbitals. 
After this factorization, $\Phi$ is replaced by $Q$ and the corresponding overlap $O\ped{T}$  by $O\ped{T}/\det(R)$ because $Q$ contains all the information about the walker $\ket{\phi}$ while $R$ only contributes to the overlap of $\ket{\phi}$. 
With importance sampling, only the information in $Q$ is relevant and $R$ can be discarded.

\section{Algorithm}
\label{sec: algorithm}

\begin{enumerate}[(1)]
\item For each walker, specify its initial state. Here we use the trial wave function $\Phi\ped{T}$ as the initial state and assign the weight $w$ and overlap $O\ped{T}$ each a value of unity.
\item \label{enumerate:BK/2}If the weight of a walker is nonzero, propagate it via $B\ped{K/2}$ as follows:
\begin{enumerate}[(a)]
\item Perform the matrix-matrix multiplication
\beq
\Phi'=B\ped{K/2}\Phi
\eeq
(recall the convention that $B\ped{K/2}$ denotes the matrix of $\hat B\ped{K/2}$) and compute the new importance function
\beq
O\ped{T}'=O\ped{T}(\phi')\,.
\eeq

(We can also work in momentum space, e.g., by using fast Fourier transforms.)
\item If $O\ped{T}'\ne 0$, update the walker, weight and $O\ped{T}$ as
\beq
\Phi \gets \Phi' \, , \quad w\gets w \, O\ped{T}'/O\ped{T}\, , \quad O\ped{T}\gets O\ped{T}'
\eeq
\end{enumerate}
\item If the walker's weight is still nonzero, propagate it via $B\ped{V}(\vv{x})$ as follows

\begin{enumerate}[(a)]
\item Compute the inverse of the overlap matrix
\beq
O\ped{inv} = \left( \Phi\ped{T}^\dagger \Phi\right)^{-1}
\eeq
\item For \emph{each} auxiliary field $x_i$, do the following:
\begin{enumerate}[(i)]
\item Compute $\widetilde{p}(x_i)$

\item Sample $x_i$ and update the weight as 
\beq
w\gets w [\widetilde{p}(x_i=+1) + \widetilde{p}(x_i=-1)]
\eeq
\item If the weight of the walker is still not zero, propagate the walker by performing the matrix multiplication
\beq
\Phi'=b\ped{V}(x_i)\, \Phi
\eeq
and then update $O\ped{T}$ and $O\ped{inv}$.
\end{enumerate}
\end{enumerate}
\item Repeat step~\ref{enumerate:BK/2}.
\item \label{enumerate:overallNormalization} Multiply the walker's weight by a normalization factor:
\beq
w\gets w \, \e^{\Delta \tau E\ped{T}}
\eeq
where $E\ped{T}$ is an adaptive estimate of the ground state energy $E_0$ and is calculated using the mixed estimator in~\eqref{eqn:mixed-estimator-expanded}.
\item Repeat steps~\ref{enumerate:BK/2} to~\ref{enumerate:overallNormalization} for all walkers in the population. This forms one step of the random walk.
\item If the population of walkers has achieved a steady-state distribution, periodically make measurements of the ground-state energy.
\item Periodically adjust the population of walkers. (See Section~\ref{sec:popControl}.)

\item Periodically re-orthonormalize the columns of the matrices $\Phi$ representing the walkers.
(See Section~\ref{sec:reorthonormalization}.)
\item Repeat this process until an adequate number of measurements have been collected.
\item Compute the final average of the energy measurements and the standard error of this average and then stop.
\end{enumerate}

\section{Package Overview and Instructions}
\label{sec: package}

\begin{figure}
\centering
\includegraphics[scale=0.23]{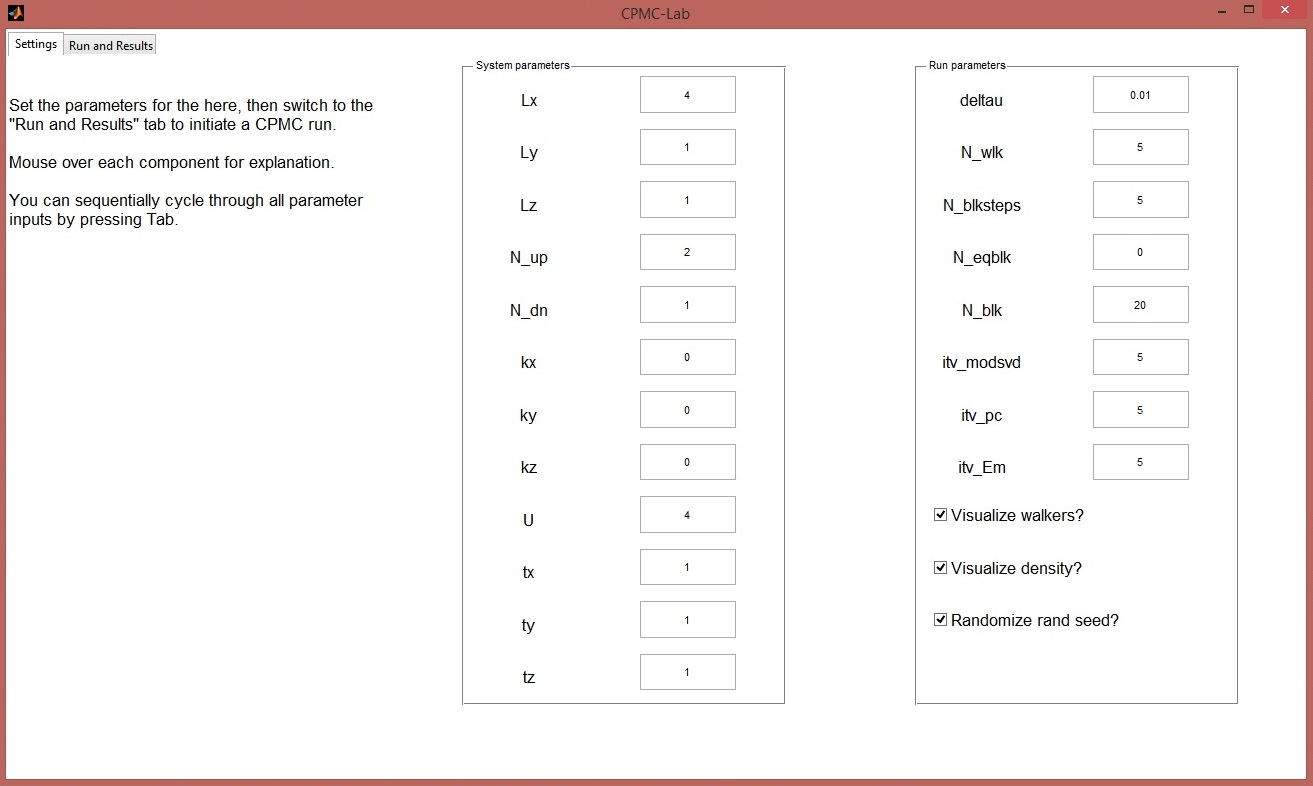}
\caption{(Color online) The GUI tab that allows the user to set all the parameters of the calculation.}
\label{fig:guiParameters}
\end{figure}

\begin{figure*}
\centering
\subfigure[With visualizations]{\includegraphics[scale=0.23]{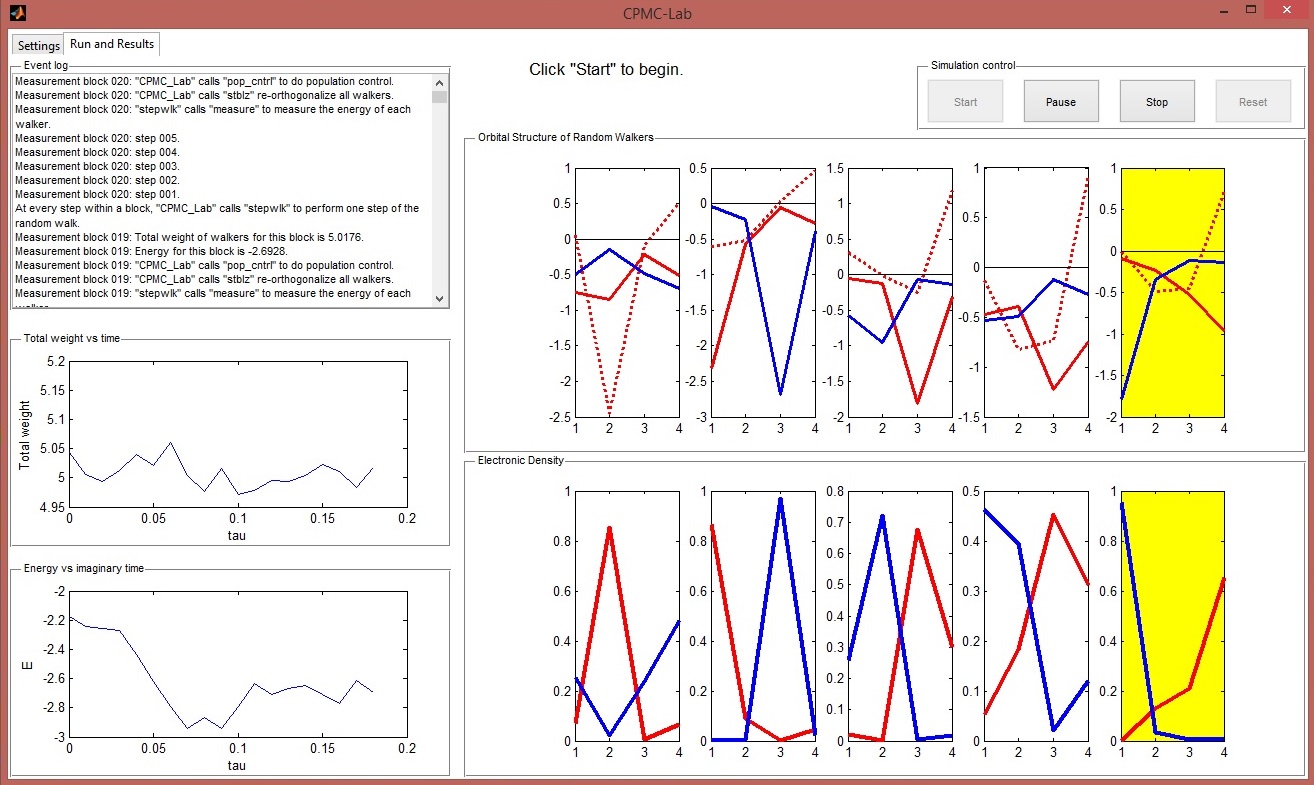}}\quad
\subfigure[Without visualizations]{\includegraphics[scale=0.23]{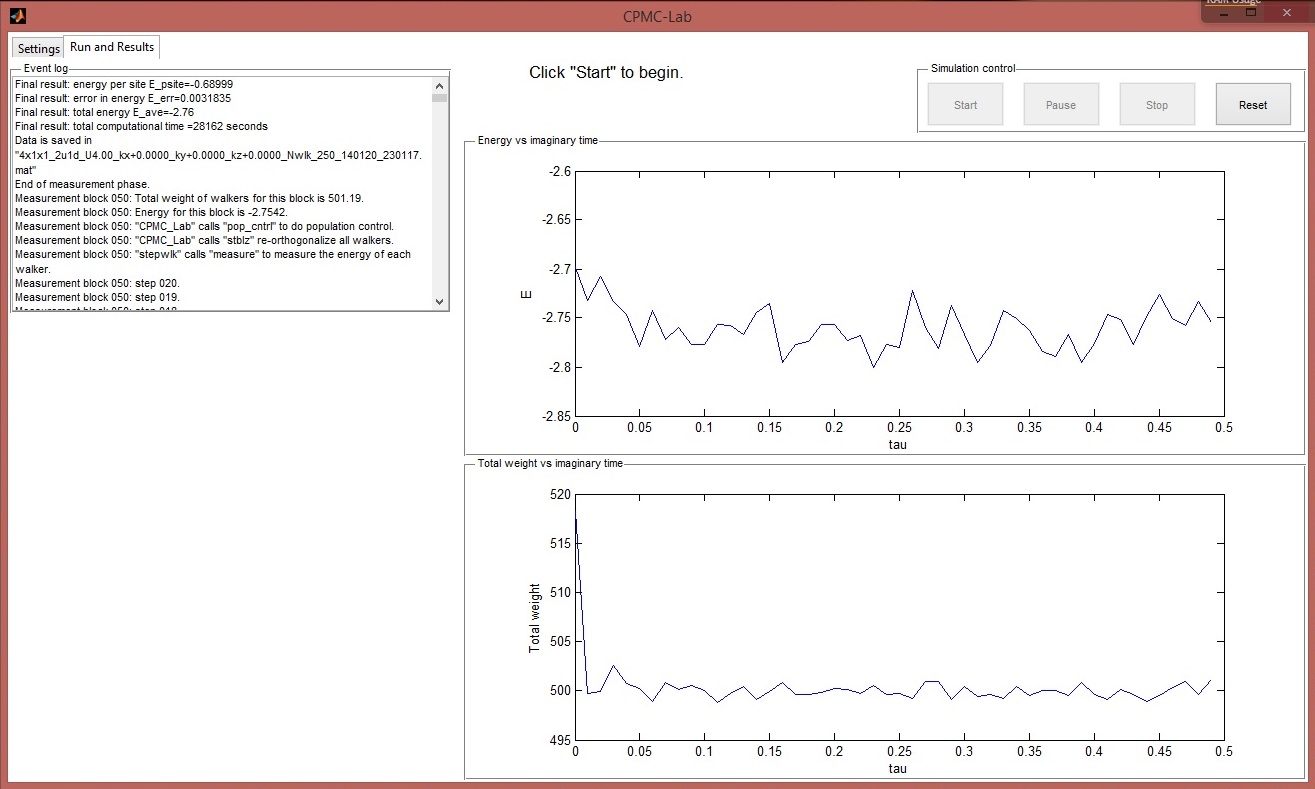}}
\caption{(Color online) The GUI for visualizing the calculation. These figures show (a) a detailed visualization mode which shows the walkers' orbitals and electronic density, with the currently propagated walker highlighted in yellow, (b) the normal run mode with the detailed visualizations turned off.}
\label{fig:guiWalkers}
\end{figure*}

The package can  be run either directly from the command line or from the GUI. The GUI requires \Matlab{} R2010b (version 7.11) and above but the non-interactive scripts can run on any version of \Matlab. The GUI is meant to minimize the initial learning process and should  be used only for small systems and short runs. Its visualization can be used to help understand how the random walkers propagate and how the auxiliary-fields build in the electron correlation. Figures~\ref{fig:guiParameters} and~\ref{fig:guiWalkers} show the program's GUI which allows users to input model parameters such as the number of sites and electrons, the twist condition, the interaction strength and the hopping amplitude. The  run parameters allow virtually any combination that the user chooses to configure the AFQMC run, including the number of walkers,  the time step $\Delta\tau$, the numbers of blocks for equilibration and  measurement, the size of the blocks, the intervals for carrying out population control and stabilization and so on. There are two tabs. The first gives the GUI for run parameters, while the second shows the progress of the calculation. Under the second tab, the total denominator of the mixed estimator and the calculated energy are monitored as a function of the imaginary time $\tau$.


The run could also be set in a more detailed visualization mode. The structure of the random walker
orbitals is illustrated by plotting the orbital coefficients, along with the electronic ``density'' $\braket{ n_{i\sigma}}$ for each spin at site $i$, defined  as $\braket{ n_{i\sigma}}=\braket{\phi_k| \hat n_{i\sigma}| \phi_k}$ for the $k$-th walker. Figure~\ref{fig:guiWalkers} shows this feature, in which a snapshot is highlighted for a four-site calculation.

\subsection{Files in the package}
This package contains ten source files for the CPMC program, one GUI program, and two scripts as samples for running the program:
\begin{description}
\item [\code{CPMC\_Lab.m}] is the main driver of the package. For every time step, it calls \code{stepwlk.m} to propagate the walkers. When appropriate, it calls \code{measure.m} to measure the energy, \code{stlbz.m} to re-orthonormalize the walkers or \code{pop\_cntrl.m} to do population control. After the end of the random walk, it calculates the final average and standard error of the energy and saves the results to a file.
\item [\code{initialization.m}] runs \code{validation.m} to conduct a basic check on the input parameters and initializes internal quantities, e.g.\ the total number of sites and electrons. It forms the free-electron trial wave function and creates the initial population of walkers.
\item [\code{validation.m}] verifies the basic validity of user inputs, as mentioned above.
\item [\code{H\_K.m}] creates the one-body kinetic Hamiltonian.
\item [\code{stepwlk.m}]  carries out one step of the random walk by calling \code{halfK.m}, \code{V.m} and \code{halfK.m} again.
\item [\code{halfK.m}] propagates a walker by $\e^{-\Delta\tau \hat K/2}$
\item [\code{V.m}] carries out importance sampling site by site to select the auxiliary fields, and propagates a walker by $\e^{-\Delta\tau \hat V}$.
\item [\code{measure.m}] computes the energy of a walker.
\item [\code{stblz.m}] orthonormalizes walkers by the modified Gram-Schmidt algorithm.
\item [\code{pop\_cntrl.m}] carries out population control by a simple-combing method.
\item [\code{sample.m}] is a script that allows users to set input parameters.
\item [\code{batchsample.m}] is a script that loops over multiple sets of parameters.
\item [\code{GUI.m}] launches the GUI of the package. It is a stand-alone file that is independent of all the other files in the package and contains all the subroutines of a QMC run.
\end{description}

\subsection{Exercises}

Below is a list of suggested exercises. They are designed as a step-by-step guide for the reader to gain a basic familiarity with the code and learn the most essential features of a CPMC or AFQMC calculation. The concepts covered in the first few exercises are universal, such as auto-correlation time, statistical errors, equilibration time, Trotter errors, population control bias, and so on.  It is essential to master them before any production runs. The other exercises are more open-ended. Their goal is to help the reader gain more insight and facilitate real applications. In the next Section, several example applications are shown.

\subsubsection{Running the sample script \code{sample.m}}
\label{subsec:exe}
The first assignment is to run a ``sample.'' To run the script \code{sample.m}, put all the files in this package directly under the current directory. Type ``sample'' (without quotes) and hit \code{Enter} to run the script. The main function \code{CPMC\_Lab} will return the values of \code{E\_ave} and \code{E\_err} to the workspace, and save more detailed data in the \code{\textsuperscript{*}.mat} file named by \code{sampledatafile}. The main function will also plot a figure of the energies from each measurement block, \code{E(i\_blk)} vs.\ $\code{i\_blk}\times\code{N\_blksteps}\times\code{deltau}$.

Save the figure via the \code{File} menu and explain its behavior. Obtain a rough estimate  for how much imaginary time is needed to equilibrate. This interval, $\tau\ped{eq}$, corresponds to the input $\code{deltau}\times\code{N\_blksteps}\times\code{N\_eqblk}$. Now modify the parameters to include an ``equilibration'' phase. Do statistical analysis on the ``measurement'' phase or do a loop over different values of \code{N\_blksteps}. We provide with the package a sample batch run script, \code{batchsample.m}, that loops over multiple sets of parameters. 
Determine the minimum length of each measurement block (\code{N\_blksteps}) necessary to obtain uncorrelated results and a reliable estimate of the statistical error. 

Now modify the parameters using the correct \code{N\_eqblk} and \code{N\_blksteps} for a few ``standard'' runs. 
Table~\ref{table:sampleRuns} gives the exact energies for  several systems ($U/t=4$) for comparison.

\begin{table*}
\centering
\begin{tabular}{ccc r@{.}l r@{.}l r@{.}l} 
\toprule
\multicolumn{2}{c}{system} &
 (\code{kx,ky}) & 
\multicolumn{2}{c}{$\braket{K}$} & 
\multicolumn{2}{c}{$\braket{V}$} & 
\multicolumn{2}{c}{$E_0$} \\ 
\hline
$2\times 1$ &$1 \uparrow 1 \downarrow$	&$(+0.0819, +0.0000)$	 &-3&54222     &   1&09963    &  -2&44260\\
$4 \times 1$	&$2 \uparrow 2 \downarrow$	& $(+0.0819, +0.0000)$	 &-3&29161      &  1&17491     &  -2&11671\\
$8 \times 1$	&$4 \uparrow 4 \downarrow$	& $(+0.0819, +0.0000)$	&-7&65166    &    3&04575   &    -4&60591\\
$2 \times 4$	&$3 \uparrow 2 \downarrow$	& $(+0.0819, -0.6052)$ &-13&7778   &     1&65680     &  -12&1210\\ 
$3 \times 4$	&$3 \uparrow 3 \downarrow$	& $(+0.02, 0.04)$ &-15&2849   &     1&29311     &  -13&9918\\ 
$4 \times 4$	&$5 \uparrow 5 \downarrow$	& $(0, 0)$ &-22&5219 &     2&94100     &  -19&58094\\
\hline
\end{tabular}
\caption{Parameters and exact results for sample runs.}
\label{table:sampleRuns}
\end{table*}

\subsubsection{Controlling the basic run parameters} \label{sec:timeStepExtrapolation}

Let us next study the behaviors of the Trotter error and population control bias:

\begin{enumerate}

\item Run the code for a few different values of $\Delta \tau$ (\code{deltau}), e.g.\ 0.025, 0.05, and 0.1, and examine the convergence of the energies as a function of \code{deltau}. Note that the number of steps in each block, \code{N\_blksteps}, the frequency of measurement, \code{itv\_em}  (and optionally \code{itv\_modsvd} and \code{itv\_pc}) should be adjusted to obtain comparable statistics.

\item Run the code for several values of population size (\code{N\_wlk}), e.g.\ 10, 20, 40, and 80, and examine the \emph{population bias} (systematic error vs.\ the population size). Note that with fewer walkers in the population, more blocks will be needed to get comparable statistical accuracy.

\end{enumerate}

\subsubsection{Calculating \code{E\_K} and \code{E\_V} separately}
\label{sec:exe3}
The program only outputs the total ground state energy. The mixed-estimate for other observables is biased, as mentioned earlier. The standard way to calculate unbiased expectation values is to use the back-propagation technique (see Refs~\cite{cpmc0K-1,PhysRevE.70.056702}). Let us calculate the kinetic energy \code{E\_K} and the potential energy \code{E\_V} in an alternative way. From the Hellman-Feynman theorem, we have
\begin{equation}
\code{E\_V} =
\Braket{\psi (U) | U \, \frac{\mathrm{d}H}{\mathrm{d}U} | \psi(U) }
= U \frac{\mathrm{d}E}{\mathrm{d}U}
\label{eqn:hellmanFeynman}
\end{equation}
Obtain $\frac{\mathrm{d}E}{\mathrm{d}U}$ by finite difference with three separate total energy calculations at $(U-\Delta U)$, $U$, and $(U+\Delta U)$ with a small $\Delta U$. Higher-order finite difference methods can also be used for more accurate results as illustrated in Section~\ref{sec:examples}. Some exact results at $\code{U}=4$ are listed in Table~\ref{table:sampleRuns} for comparison. For the two-dimensional system in the last three rows, obtain a result with sufficiently small statistical error bars to examine the systematic error from the constrained path approximation.

\subsubsection{The Hydrogen molecule}
Let us study the system of two interacting fermions on two sites: $\code{N\_up}=1$, $\code{N\_dn} = 1$, $\code{Lx} = 2$ and $\code{Ly} = 1$. We set $\code{tx} = 1$ and $\code{kx} = 0$ (\code{ty} and \code{ky} will be ignored when $\code{Ly} = 1$). Study the properties of the system as the interaction strength \code{U} is varied (while keeping $t=1$). This can be viewed as a crude model (minimal basis) for breaking the bond in a H$_2$ molecule. As the distance between the two protons  increases, the interaction strength $U/t$ increases.

\begin{enumerate}
\item Run the QMC code at different values of \code{U} and compare your results with the exact solution:
\begin{equation}
E = \frac {1}{2} \left( U-\sqrt{U^2+64} \, \right)
\end{equation}

\item  Plot \code{E\_K} and \code{E\_V} vs.\ \code{U}. Explain their behaviors.

\item Obtain the double occupancy $\braket{ n_{1 \uparrow} n_{1 \downarrow} }$ (see Section~\ref{sec:exe3}). From it, derive the correlation function $\braket{ n_{1 \uparrow} n_{2 \downarrow} }$.  Explain its behavior  vs.\ \code{U}. 
\end{enumerate}

\subsubsection{Ground-state energy of a chain}
Let us study the half-filled Hubbard model in one dimension and how the energy converges with respect to system size. Run the program  for a series of lattice sizes (e.g.\ $2 \times 1$, $4 \times 1$, $6 \times 1$, $8 \times 1$, \dots), each averaging over a set of random twist angles \code{kx}. Plot the energy per site vs.\ $\code{1/L}^2$, and examine its convergence behavior. (In Section~\ref{sec:examples}  a detailed example is given.)
	
\subsubsection{Addition to the program: other correlation functions}
 Program in the mixed estimator of some observables, e.g.\ the one-body density matrix 
 $\braket{ c^\dagger_{i \sigma} c_{j \sigma} }$, 
 the spin-spin correlation function $\braket{ S_i S_j }$ (where $S_i=n_{i \uparrow}-n_{i \downarrow}$), 
 and the charge-charge correlation function $\braket{ (n_{i \uparrow}+n_{i \downarrow})(n_{j \uparrow}+n_{j\downarrow}})$. Calculate the mixed estimate for  \code{E\_K} and \code{E\_V} and compare with the results from Sec.~\ref{sec:exe3}.

\section{Computational Speed} \label{sec:speed}

As mentioned in Section~\ref{sec: intro}, a major drawback of the \Matlab{} package is that it is significantly slower than a standard production code written in \textsc{FORTRAN} or \textsc{C}. This is outweighed by the advantages in pedagogical value and in providing the clearest algorithmic foundation. From this foundation, the users could build a CPMC or phase-free AFQMC code tailored toward their own applications. As illustrated in the next section, significant applications can be carried with the present \Matlab{} code as is. 
 
Here we give some rough comparisons between the \Matlab{} code and a production code in  \textsc{FORTRAN}. To provide an idea of the timing difference, we describe two examples. For the calculation at $U=4$ in  a $4\times 4$ lattice with 5 spin-$\uparrow$ and 5 spin-$\downarrow$ electrons (the same number of lattice sites as Figure~\ref{fig:energyVsU}), the \textsc{FORTRAN} code takes 1 minute to run on an Intel Core i7-2600 3.40 GHz processor, compared to 32 minutes for the \Matlab{} code. Scaling up the system size to a $128\times 1$ lattice with 65 spin-$\uparrow$ and 63 and spin-$\downarrow$ electrons (the largest system in Figure~\ref{fig:spinGap}), the \textsc{FORTRAN} code requires 186 minutes while the  \Matlab{} code takes 460 minutes. The  parameters for both run are $\code{deltau}=0.01$, $\code{N\_wlk}=1000$, $\code{N\_blksteps}=40$, $\code{N\_eqblk}=10$, $\code{N\_blk}=50$, $\code{itv\_modsvd}=5$, $\code{itv\_pc}=40$ and $\code{itv\_Em}=40$.

Using OpenMP, \Matlab{} can automatically speed up computations in a multi-core environment. Furthermore, \Matlab{} users with \Matlab's Parallel Computing Toolbox installed can easily parallelize the code by distributing the propagation of individual walkers over multiple processor cores.
This is done by
\begin{inparaenum}[1)]
\item changing the main \code{for} loop in \code{stepwlk.m} into a \code{parfor} loop and
\item opening a pool of $n$ parallel \Matlab{} workers with the command \lstinline[basicstyle=\small\ttfamily]{matlabpool('open',n)} before the equilibration phase begins in \code{CPMC\_Lab.m}.
\end{inparaenum} 

The computational cost of the CPMC and phase-free AFQMC methods scales algebraically, roughly as the third 
 power of system size. (Different pieces of the code scale as different combinations of $N$ and $M$.) 
The memory required to run \code{CPMC-Lab} is proportional to the product of (the basis size) $\times$ (the number of electrons) $\times$ (the number of random walkers). 
The random walkers are only loosely coupled. The approach  
is ideally suited for a distributed massively parallel environment \cite{2008-SCIDAC-EndStation}.

\section{Illustrative Results} \label{sec:examples}

\begin{figure}
\centering
\subfigure[ Total energy]{\includegraphics[scale=1]{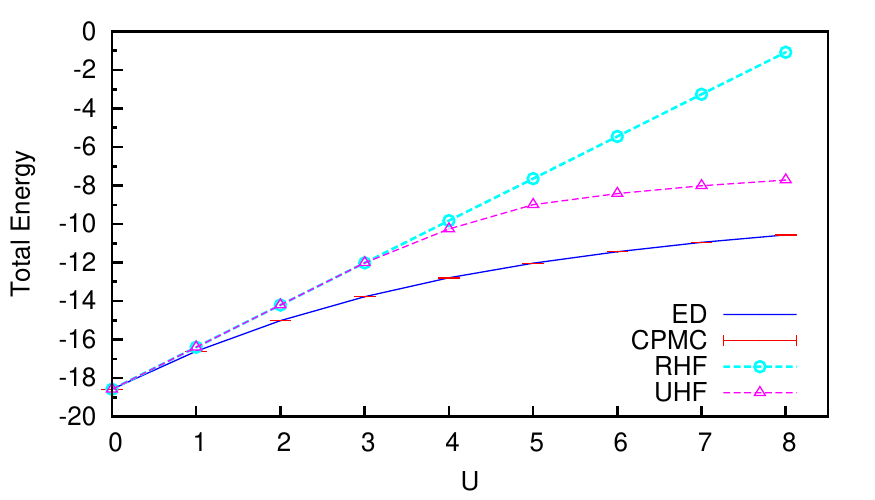}}\quad
\subfigure[ Potential energy (inset: double occupancy $\braket{n_{\uparrow}n_{\downarrow}}$)]{\includegraphics[scale=1]{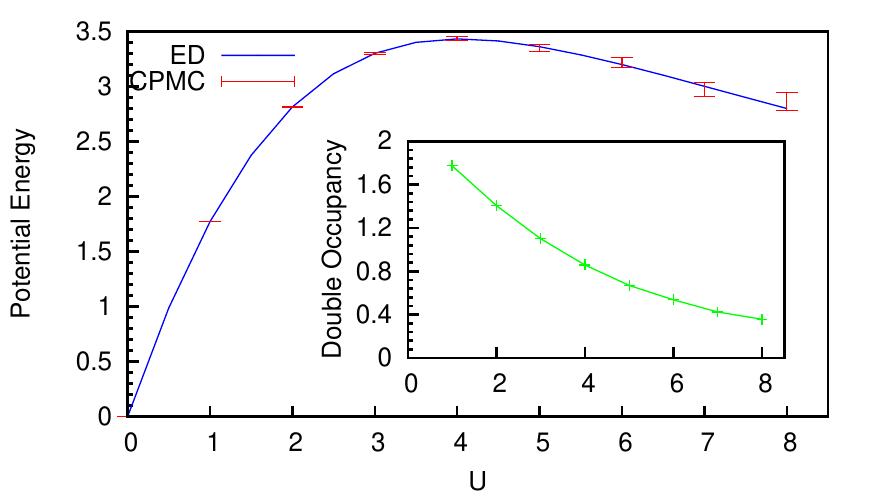}}\\
\subfigure[ Kinetic energy]{\includegraphics[scale=1]{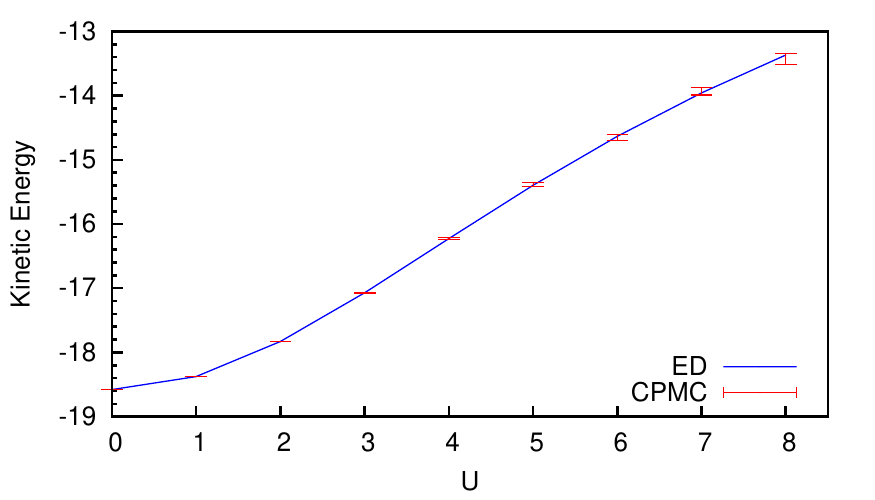}}
\caption{(Color online) The total, potential and kinetic energies versus the interaction strength $U$ for a 16-site ring with 
$5$ spin-$\uparrow$ and $7$ spin-$\downarrow$ electrons. CPMC results (red error bars) are 
compared with exact diagonalization (ED) (blue solid curves). Restricted (cyan) and unrestricted (magenta) Hartree-Fock results have also been drawn for comparison in panel (a). The inset in panel (b) shows the 
double occupancy $\braket{n_{\uparrow}n_{\downarrow}}$.}
\label{fig:energyVsU}
\end{figure}

Figure~\ref{fig:energyVsU} compares energy calculations by \code{CPMC-Lab} against exact diagonalization (ED) results for a one-dimensional 16-site Hubbard model with $5$ spin-$\uparrow$ and $7$ spin-$\downarrow$ electrons. The parameters of the run are $\code{N\_wlk}=5000$, $\code{deltau}=0.01$, $\code{N\_blksteps}=40$, $\code{N\_blk}=150$, $\code{N\_eqblk}=30$, $\code{itv\_pc}=5$, $\code{itv\_Em}=40$ and $\code{itv\_modsvd}=1$. The potential energy is obtained by the Hellman-Feynman method of  Eq.~\eqref{eqn:hellmanFeynman}, where the derivative $\frac{\mathrm{d}E}{\mathrm{d}U}$ is calculated using the five-point stencil. The kinetic energy is simply the difference between the total and potential energy. 

In one-dimension, the CPMC method is exact. As seen in the figure, the agreement between CPMC and ED results is excellent. As the interaction strength $U$ increases, the kinetic energy also increases because more electrons are excited to occupy higher single-particle levels. The potential energy is non-monotonic as a consequence of two opposing tendencies. Double occupancy $\braket{n_{\uparrow}n_{\downarrow}}$ (shown in the inset) is rapidly reduced as the interaction is increased. 
On the other hand, the growing value of $U$  increases the potential energy linearly. 

\begin{figure}
\centering
\includegraphics[scale=1]{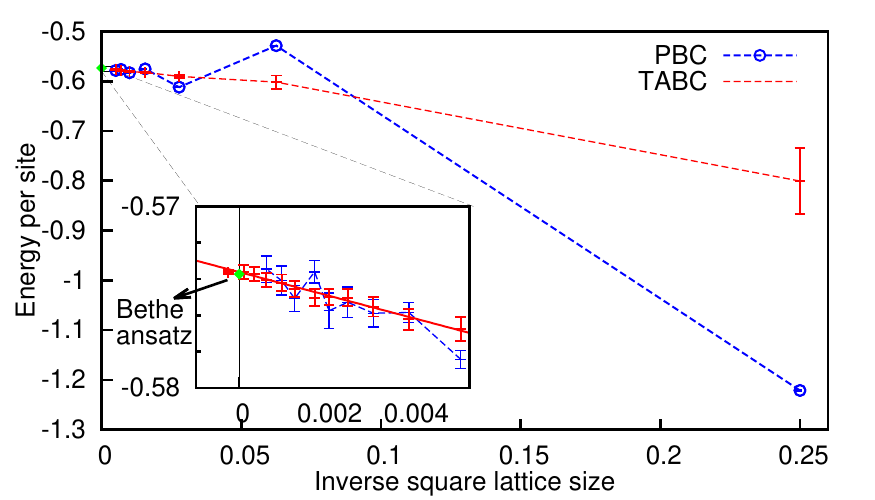}
\caption{(Color online) Ground-state energy per site vs.~inverse square lattice size for half-filled one-dimensional lattices at $U=4$. Results from PBC (blue) and TABC (red) are shown. The lattice size in PBC ranges from 2 to 40 and those in TABC are from 2 to 128. The red solid line in the inset is the best fit of the TABC data (for lattice sizes 4 to 128) while the dashed blue line is to guide the eye. The inset shows a closeup view (from 12 to 128 sites) of the convergence to the thermodynamic limit. The exact result is shown in green for comparison. For clarity, the error bar of the intercept in the inset is offset to the left of the green symbol for exact result.}
\label{fig:energyVsL}
\end{figure}

We next compute ground-state properties of the one-dimensional Hubbard model in the thermodynamic limit. Since our QMC calculations are performed in finite-sized supercells, it is important to reduce finite-size effects and obtain better convergence as $M\to \infty$. In Figure~\ref{fig:energyVsL}, we show the calculated ground-state energy per site versus the supercell size at half-filling for $U=4$. We compare two sets of data, one with PBC ($\Gamma$-point) and the other with twist-averaging boundary condition (TABC) as discussed in Section~\ref{sec:phaseProblem}. In the TABC runs, each data point is obtained by averaging over random samples of the twist angle $\bm{\Theta}$. That is, the \code{kx} values are chosen randomly from the interval $(-1,1]$, corresponding to $\theta_x\in (-\pi,\pi]$. The number of  \code{kx} points for each lattice size is chosen to keep the product $(\code{lattice size})\times(\code{number of kx values})$  roughly constant ($\sim 80$ in this case) while maintaining a minimum of 4 points. 

In the PBC runs, we added a small twist angle \code{kx} whenever the electron configuration is open-shell in order to break the degeneracies in the single-particle energy levels (see Section~\ref{sec:phaseProblem}). The run parameters for this and subsequent results are identical to those used in Figure~\ref{fig:energyVsU} except for $\code{N\_wlk}=1000$, $\code{itv\_pc}=40$ and $\code{itv\_modsvd}=5$. 

The PBC data is seen to exhibit large finite-size effects, with a zig zag pattern reflecting differing trends for closed-shell and open-shell systems. We reiterate that this behavior is not from any numerical problem; rather it is the nature of the exact ground-state eigenvalue of the Hamiltonian for these finite supercells under PBC.  The TABC data, on the other hand, is smooth and monotonic. As the supercell size increases, the two sets of data approach each other, and converge to the same 
limit. The best fit of the TABC data to a straight line is shown in the figure. The fit is given by $E_0/M= - 0.57361-0.63192/M^2$, leading to a thermodynamic value of $- 0.5736(1)$, to be compared to the exact result of $-0.573729$ from the Bethe ansatz~\cite{LiebWu}. 

We next compute the spin and charge excitation gaps. The \emph{spin gap} is defined as the energy difference between the present system and that with one spin flipped:
\beq
\Delta\ped{s}=E_0\left(N_\uparrow+1,N_\downarrow-1 \right) - E_0\left(N_\uparrow,N_\downarrow \right)\,, \label{eqn:spinGap}
\eeq
where $E_0\left(N_\uparrow,N_\downarrow\right)$ is the total ground-state energy for a finite supercell with 
$N_\uparrow$ spin-$\uparrow$ and $N_\downarrow$ spin-$\downarrow$  electrons.

\begin{figure}
\centering
\includegraphics[scale=1]{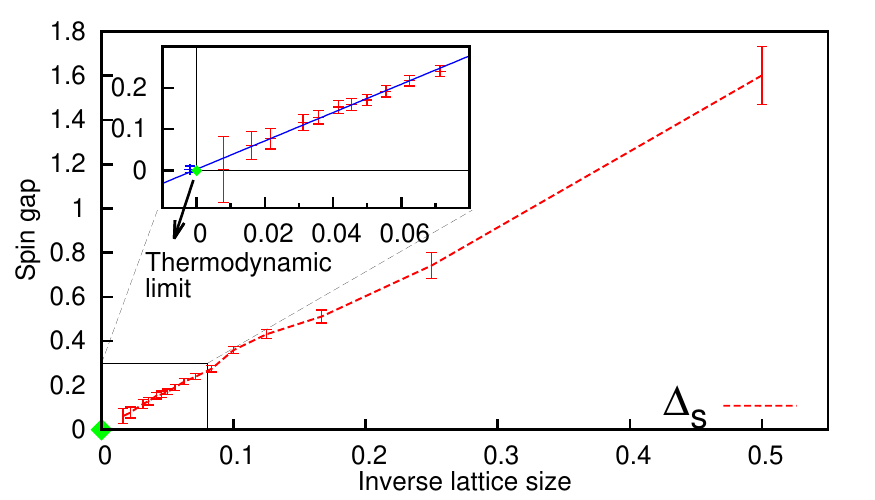}
\caption{(Color online) The calculated spin gap vs. inverse lattice size for one-dimensional Hubbard
model at half-filling, $U=4$. The red dashed line is drawn to guide the eye.  The inset shows a closeup view of the convergence from 12 to 128 sites. The blue solid line is a linear fit of the data for lattice sizes 8 to 128. For clarity, the error bar of the intercept in the inset is offset to the left of the vertical axis at $1/M=0$.}
\label{fig:spinGap}
\end{figure}

Figure~\ref{fig:spinGap} shows the result of the spin gap for the one-dimensional Hubbard model at half-filling, for $U=4$, as a function of inverse lattice size. TABC is used in these calculations, choosing a random set of twist values to calculate each total energy in Eq.~\eqref{eqn:spinGap}. The parameters of the run are identical to those in Figure~\ref{fig:energyVsU}.

As the lattice size increases, the calculated spin gap converges smoothly. A linear fit to all the data yields an asymptotic value of $\Delta\ped{s}=0.0036(80)$, consistent with the Haldane conjecture of a zero spin gap~\cite{Haldane1983464,Haldane1983}.

\begin{figure}
\centering
\includegraphics[scale=1]{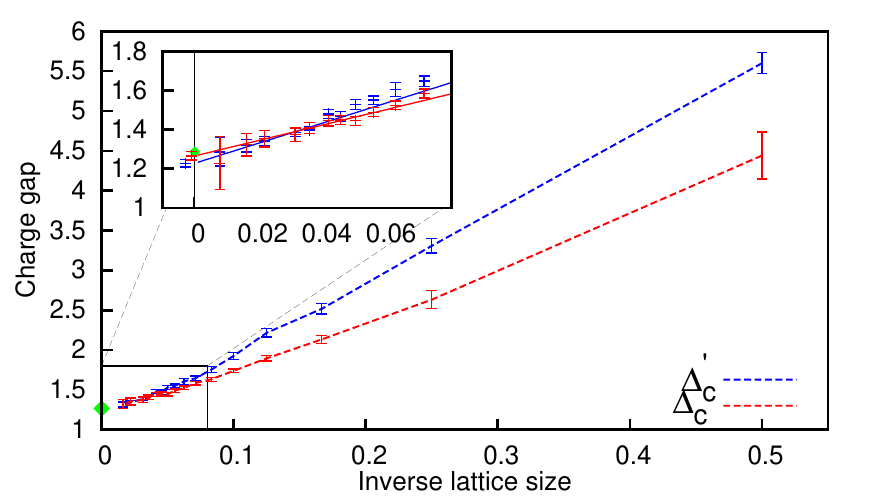}
\caption{(Color online) The calculated charge gap vs. inverse lattice size for the one-dimensional Hubbard model at half-filling, $U=4$. The lattice size ranges from 2 to 128. The two sets of data are obtained according to $\Delta\ped{c}'$ (blue) and $\Delta\ped{c}$ (red), respectively. The dashed curves in the main figure are drawn to guide the eye. The inset shows a closeup view at larger supercell sizes. The red and blue solid lines represent linear fits to the corresponding subsets of data with lattice sizes 16 to 128. The exact result of $1.28673$ in the thermodynamic limit is indicated in green for comparison. For clarity, in the inset, the error bars for the two intercepts are offset to the left of the green symbol for exact result.}
\label{fig:chargeGap}
\end{figure}

The \emph{charge gap} is defined by the addition and removal energy:
\beq
\Delta\ped{c}=E_0\left(N_\uparrow+1,N_\downarrow \right) + E_0\left(N_\uparrow-1,N_\downarrow \right)
- 2\,E_0\left(N_\uparrow,N_\downarrow \right)\,, \label{eqn:chargeGap2}
\eeq
where in the present case $N_\uparrow=N_\downarrow=M/2$. We also calculate, alternatively 
\beq
\Delta\ped{c}'=E_0\left(N_\uparrow+1,N_\downarrow +1\right) 
- E_0\left(N_\uparrow,N_\downarrow \right)-U\,, \label{eqn:chargeGap1}
\eeq
which we expect to give a consistent result based on the analytic result relating the chemical potentials
to $U$~\cite{LiebWu}. The results of the calculated charge gap at $U=4$ are shown in Figure~\ref{fig:chargeGap}. 

As the lattice size increases from 2 to 128 (right to left), both types of charge gap converge toward a finite value. A linear fit to the data of $\Delta\ped{c}$ yields a charge gap of $1.266(21)$ in the thermodynamic limit, consistent with the exact analytic result of $1.28673$~\cite{LiebWu}. 

\section{Summary} \label{sec:summary}

In this paper we described \code{CPMC-Lab}, an open-source \Matlab{} program for studying the many-body ground state of Hubbard-like models in any dimension, and for learning  the constrained-path and phase-free auxiliary-field Monte Carlo methods. The package illustrates the constrained-path Monte Carlo method, with a graphical interface. The ground-state energy is calculated using importance sampling and implementing the algorithmic details of a total energy calculation. This tool allows users to experiment with various model and run parameters and visualize the results. It  provides a direct and interactive environment to learn the method and study the code with minimal overhead for setup. 
Furthermore, it provides a foundation and template for building a CPMC or phase-free AFQMC calculation for essentially any interacting many-fermion system with two-body interactions, including \emph{ab initio} calculations in molecules and solids.

\section{Acknowledgment}
We gratefully acknowledge support by the US National Science Foundation Grant nos.\ DMR-1006217 and DMR-1409510 (H.N., J.X., and S.Z.), by the Department of Energy under 
grant no.~DE-SC0008627 (H.S.~and S.Z.) and by Reed College under the Summer Experience Award (H.N.). Computing support was provided by a DOE INCITE Award and from 
 William and Mary's SciClone cluster.

 \appendix

\section{Some useful  \Matlab{} commands} \label{sec:matlabCommands}
\lstset{basicstyle=\small\ttfamily}
\begin{itemize}
\item Tab completion is available in the \Matlab{} command window for the names of functions and scripts (either built-in, on the search path or in the current directory) and variables in the current workspace. 
\item Ending a command with semicolon suppresses its output.
\item To display the value of the variable \code{variablename}: 
\begin{lstlisting}
variablename
\end{lstlisting}
\item To display a brief description and the syntax for \code{functionname} in the command window:
\begin{lstlisting}
help functionname
\end{lstlisting}
\item To call a function:
\begin{lstlisting}
[out1,out2,...]=myfunc(in1,in2,...)
\end{lstlisting}
\item To run the script \code{scriptname}:
\begin{lstlisting}
scriptname
\end{lstlisting}
\item To set display format to \code{long} (15 decimal places) instead of the default \code{short} (4 decimal places):
\begin{lstlisting}
format long
\end{lstlisting}
\item To load the variables from a *.mat file into the workspace:
\begin{lstlisting}
load filename.mat
\end{lstlisting}
or double click the *.mat file or select \code{Import Data} from the workspace menu
\item To remove all variables from the workspace:
\begin{lstlisting}
clear
\end{lstlisting}
\item To generate an $m\times n$ matrix containing pseudorandom values
drawn from the standard uniform distribution on  $(0, 1)$:
\begin{lstlisting}
rand (m, n)
\end{lstlisting}
\item To create a new individual figure window on the screen:
\begin{lstlisting}
figure
\end{lstlisting}
\item To close all the figure windows:
\begin{lstlisting}
close all
\end{lstlisting}
\item To plot each column in the (real) matrix \code{Ydata} versus the index of each value:
\begin{lstlisting}
plot(Ydata)
\end{lstlisting}
\item To plot each column in the matrix \code{Ydata} versus \code{Xdata}:
\begin{lstlisting}
plot(Xdata,Ydata)
\end{lstlisting}
\item To plot each vector in \code{Yn} versus the corresponding vector \code{Xn} on the same axes:
\begin{lstlisting}
plot(X1,Y1,X2,Y2,...,Xn,Yn)
\end{lstlisting}
\item To plot \code {Ydata} versus \code{Xdata} with symmetric error bars 2*\code{Err(i)} long:
\begin{lstlisting}
errorbar(Xdata,Ydata,Err)
\end{lstlisting}
\item To hold the plot so that subsequent plotting commands add to the existing graph instead of replacing it:
\begin{lstlisting}
hold all
\end{lstlisting}
\item To put a title string at the top-center of the current axes:
\begin{lstlisting}
title('string')
\end{lstlisting}
\item To label the x-axis:
\begin{lstlisting}
xlabel('string') 
\end{lstlisting}
and similarly for the y-axis
\end{itemize}

\bibliographystyle{apsrev4-1}
\bibliography{CPMCLab1}

\end{document}